\documentclass[twocolumn,usenatbib,emulateapj]{aastex63}
\graphicspath{{./}{figures/}}

\usepackage{natbib}
\usepackage{amsmath}
\usepackage{graphicx}
\usepackage[T1]{fontenc}
\usepackage{booktabs}
\usepackage{lipsum}

\usepackage{txfonts} 
\usepackage[figure,figure*]{hypcap} 
\usepackage[utf8]{inputenc}

\newcommand{\vwind}{v_{\rm wind}}
\newcommand{\vesc}{v_{\rm esc}}
\newcommand{\tfinal}{t_{\rm final}}
\newcommand{\fism}{f_{\rm ISM}}   
\newcommand{\ficm}{f_{\rm ICM}}  
\newcommand{\Mism}{M_{\rm ISM}}   
\newcommand{\Micm}{M_{\rm ICM}}  

\begin{document}
\title{It's Cloud's Illusions I Recall:\\ Mixing Drives the Acceleration of Clouds from Ram Pressure Stripped Galaxies}%
\correspondingauthor{Stephanie Tonnesen}
\email{stonnesen@flatironinstitute.org}

\author[0000-0002-8710-9206]{Stephanie Tonnesen}
\affiliation{Center for Computational Astrophysics, Flatiron Institute, 162 5th Ave, New York, NY 10010, USA}
\author[0000-0003-2630-9228]{Greg L. Bryan}
\affiliation{Department of Astronomy, Columbia University, 550 W 120th Street, New York, NY 10027, USA}
\affiliation{Center for Computational Astrophysics, Flatiron Institute, 162 5th Ave, New York, NY 10010, USA}


\begin{abstract}

Ram Pressure Stripping can remove gas from satellite galaxies in clusters via a direct interaction between the intracluster medium (ICM) and the interstellar medium.  This interaction is generally thought of as a contact force per area, however we point out that these gases must interact in a hydrodynamic fashion, and argue that this will lead to mixing of the galactic gas with the ICM wind.  We develop an analytic framework for how mixing is related to the acceleration of stripped gas from a satellite galaxy.  We then test this model using three "wind-tunnel" simulations of Milky Way-like galaxies interacting with a moving ICM, and find excellent ageement with predictions using the analytic framework. Focusing on the dense clumps in the stripped tails, we find that they are nearly uniformly mixed with the ICM, indicating that all gas in the tail mixes with the surroundings, and dense clumps are not separate entities to be modeled differently than diffuse gas.  We find that while mixing drives acceleration of stripped gas, the density and velocity of the surrounding wind will determine whether the mixing results in the heating of stripped gas into the ICM, or the cooling of the ICM into dense clouds.

\end{abstract}


\section{Introduction}

As satellite galaxies orbit within a cluster, their interstellar medium (ISM) may interact directly with the intracluster medium (ICM), the hot halo of gas bound by the cluster gravitational potential. One of the most commonly invoked ISM-ICM interactions is ram pressure stripping, initially outlined by Gunn \& Gott (1972; GG72).  This work introduced the straightforward competition between a contact force per unit area exerted by the ICM ($P_{ram}=\rho v^2$) competing with a gravitational restoring force per area from the galaxy ($2\pi G \Sigma_* \Sigma_g$), which remains a commonly used measure today for determining whether a galaxy is likely to be ram pressure stripped and to what radius (e.g. Jaffe et al. 2018; 2019).  Indeed, simulations have verified that using this equation to determine the stripping radius (the radius outside of which gas will be removed) is reasonably accurate (Roediger \& Bruggen 2007; Tonnesen \& Bryan 2009).

Thinking of ram pressure as physically pushing gas out of a galaxy is extremely useful, and can explain several observations of ram pressure stripped galaxies.  For example, as predicted by the GG72 picture, gas has been observed to be stripped from the outside-in (Gullieuszik et al. 2017; Pappalardo
et al. 2010; Abramson et al. 2011; Merluzzi et al. 2016; Fossati
et al. 2018; Cramer et al. 2019).  In addition, high-density gas with a higher $\Sigma_g$ may survive in the disk while lower-density gas is stripped.  This has been observed in molecular clouds surviving in the disk of NGC 4402 (Crowl et al. 2005), and predicted in simulations (Tonnesen \& Bryan 2009).  This differential stripping may be followed by differential acceleration, in which low density gas is pushed to higher velocities than more dense gas (Tonnesen \& Bryan 2010; Jachym et al. 2017).  

However, thinking of the boundary between the ICM and ISM as impermeable denies the nature of the hydrodynamic interaction between the two fluids.  This boundary is often unstable to a variety of hydrodynamic or hydromagnetic instabilities, which can drive mixing and the development of intermediate temperature and density gas at constant pressure (Begelman \& Fabian 1990).  For example, Rayleigh-Taylor and Kelvin-Helmholtz instabilities may completely destroy a cold clump subjected to a hot wind (Chandrasekhar 1961; Agertz et al. 2007).  In addition, non-ideal processes may drive mixing: heat conduction can evaporate cold gas into the ICM (Cowie \& Songaila 1977; Cowie \& McKee 1977).  Shock heating of the cool gas may allow these mixing processes to act more quickly in stripped tails.  In addition to heating cold gas, radiative cooling can lead to entrainment of the hot ICM onto cold gas and mixing in surviving cold gas clouds (Klein et al. 1994; Mellema et al. 2002; Scannapieco \& Br{\"u}ggen 2015; Gronke \& Oh 2018, 2019).

In addition, if we apply the ram pressure model simply requiring ram pressure to overcome the gravitational restoring force per area on small scales, then it becomes clear that dense clouds, with their large restoring forces and low surface areas, should not be susceptible to stripping, leaving dense molecular clouds impervious to ram pressure. Nevertheless, both observations (Moretti et al. 2018, 2020; Sivanandam et al. 2010; Jachym et al. 2014, 2019; Cramer et al. 2019) and simulations (e.g., Tonnesen \& Bryan 2010) demonstrate that dense gas is present in the stripped tail.\footnote{The ram-pressure literature sometimes differentiates between ram-pressure and viscous-stripping, with the implication that a non-ideal viscous force is responsible for removing denser material; however, a coherent physical picture of viscous stripping has not been developed to date.} This puzzle has been remarked on in other contexts -- for example, Thompson et al. (2016) pointed out (in the context of cold clumps in galactic winds) that clouds are destroyed more rapidly than they are accelerated to the surrounding wind speed, at least in the absence of strong radiative cooling, a statement which has been verified by high-resolution simulations (Schneider \& Robertson 2018).

 There is also observational evidence of gas mixing in stripped tails.  When comparing the expected gas mass of ram pressure stripped galaxies to the mass of gas that is observed in the disk and tail, observers often find lower masses (e.g. Vollmer \& Huchtmeier 2007; Ramatsoku et al. 2019; 2020), possibly indicating stripped gas is mixing into the ICM.  Observations also find that the metallicity in the stripped gas is often between that of the ISM and ICM (Fossati et al. 2016; Gullieuszik et al. 2017; Bellhouse et al. 2019).  

In this paper, we propose a model of ram-pressure that is driven by mixing processes, rather than a traditional contact force exerted between by the low-density wind and the high-density cloud. In particular, we write down a few straightforward relations based on this supposition (and some simple ideas about energy conservation) which describe how galactic gas is removed and accelerated from a disk in a ram-pressure-stripping wind.  We then compare it to the acceleration of gas from galaxies in wind-tunnel ram-pressure-stripping simulations.  In our comparison we examine all of the stripped gas, but also focus in one section on dense clouds that therefore have longer conduction, viscous stripping, and Kelvin-Helmholtz timescales.  Thus we are focusing on the gas that is the most likely to remain unmixed with the ICM in a classic ``pushing" scenario.

This idea, that the acceleration of cold clouds by hot flows is fundamentally driven by mixing, has received some additional support by two recent works, one examining the inflow of cold gas on to a galactic disk (Melso, Bryan \& Li 2019), and another exploring galactic outflows at high resolution (Schneider et al. 2020).

We begin by introducing our mixing model for gas stripping and acceleration in Section \ref{sec:mixing}.  In order to test the predictions from our model, in Section \ref{sec:method} we describe our simulations (Section \ref{sec:sim}) and how we identify and measure properties of clouds in the tail (Section \ref{sec:selection}).  We then test our model predictions through comparisons with all the gas behind the simulated galaxies as well as with denser clouds (Section \ref{sec:comparison}).  In Section \ref{sec:survival} we discuss the survival of the clouds in our simulated stripped tails.  We use cloud properties to determine to what extent the surrounding ICM influences stripped gas in Section \ref{sec:ICMinfluence}. We discuss the implications of our results in Section \ref{sec:discussion}, focusing on caveats in Section \ref{sec:caveats}. Finally, in Section \ref{sec:conclusion} we summarize our conclusions.

\section{Gas Acceleration via Mixing}\label{sec:mixing}

In this section we introduce a simple mixing model to try to describe the properties of a cloud of mass $\delta M$ in the downstream wake. The cloud is assumed to be a coherent identity that is identified as a single (cold) object in the wake (to inform comparison with the simulations below). This cloud can be thought of being composed of gas with two sources: (1) the cold ISM, at rest with respect to the galaxy (our reference frame), and (2) the hot ICM wind with velocity $v_{\rm wind}$. We denote the masses of those two sources contributing to the cloud as $\Mism$ and $\Micm$, respectively. Note that the gas from these two sources may not come from contiguous regions in either the ISM or the ICM. Mass conservation implies
  $\Mism + \Micm = M_{\rm total}$.
Defining fraction mass contributions $\fism = \Mism / M_{\rm total}$ and $\ficm = \Micm / M_{\rm total}$, we can rewrite this as
\begin{equation}
\fism + \ficm = 1. 
\end{equation}

Determining the resulting velocity of the gas which ends up in the cloud is, of course, more challenging (which we assume to be well mixed, a question directly addressed in the simulation investigations explored later in the paper).  However, our overriding assumption is that acceleration of the cloud is done through the process of mixing, and that mixing of the hot (wind) phase into the cold cloud leads to a gain of mass, momentum and energy.

  Here, we explore two simple models, one based on momentum conservation neglecting the potential of the host galaxy, and a second approach based on energy conservation.

If we ignore any external forces on the gas during the mixing process (as well as any gravitational torques), then a simple application of momentum conservation implies that the cloud velocity would be given by
\begin{equation}
v_m = \ficm v_{\rm wind}
\end{equation}
where we use subscript $m$ to denote the momentum conservation estimate (see also Gronke \& Oh 2018; Schneider et al. 2020).  In principle, we could include the gravitational forces, but this would require a knowledge of the trajectory of the gas elements throughout it's evolution, which requires more assumptions.  In addition, $v_{\rm wind}$ should more accurately be the velocity after being processed through the bow shock (if present), which reduces the momentum of the gas by a factor which depends on the Mach number $\mathcal{M}=v_{\rm wind}/c_s$ (where $c_s$ is the sound speed in the hot gas).  For low Mach numbers and subsonic flow this is negligible, and even for the highest velocity case we explore, we find that the ICM flow velocity in the vicinity of active cloud entrainment is only mildly reduced from $\vwind$.

The other extreme is to assume a form of energy conservation. On first blush, this seems problematic, both because radiative loses are clearly important and also because of work done on this gas during its evolution, however we can make simple assumptions as to the amount of energy lost due to radiative cooling as well as assume that no work is done during the mixing. We express this with:
\begin{equation}
\begin{split}
\ficm v_{\rm wind}^2 \left(1 + \frac{2}{(\gamma-1)\mathcal{M}^2} \right) & - \fism v_{\rm esc}^2  = \\
& v_e^2 + \ficm \chi \left( \frac{2}{(\gamma -1) \mathcal{M}^2} \right) v_{\rm wind}^2
\end{split}
\end{equation}
The left-hand side is the `before' state; the first term represents the total enthalpy plus kinetic energy of the wind gas, while the gas in the galaxy only contributes a term due to its gravitational potential energy, expressed in terms of the escape velocity ($v_{\rm esc}$).  We neglect any gravitational contribution for the ICM wind as well as for the cloud after stripping, under the assumption that they are sufficiently far from the galaxy.  In the `after' state on the right-hand-side, we assume that the gas is cold and so only has a kinetic component (in this case, we denote the final cloud velocity as $v_e$ to emphasize the energy formulation behind the estimate).  We account for the radiative losses with the second expression on the right-hand-side.  Our ignorance as to the amount of energy radiated is expressed in terms of the ratio to the incoming enthalpy, such that $\chi = 1$ is the minimum energy loss according to our assumption that the gas is cold in the `after' state.

Making the $\chi = 1$ assumption simplifies this to
\begin{equation}
\begin{split}
v_e = \left( \ficm v_{\rm wind}^2 - \fism v_{\rm esc}^2   \right)^{1/2}& = \\
& \left( \ficm \left(v_{\rm wind}^2 + v_{\rm esc}^2\right) - v_{\rm esc}^2   \right)^{1/2}
\end{split}
\end{equation}
This implies a minimum amount of mixing from the ICM wind to launch a cloud, since $v_e$ only becomes real and positive when $f_{\rm ICM} > f_{\rm ICM,{\rm crit}} = v_{\rm esc}^2/(v_{\rm wind}^2 + v_{\rm esc}^2)$.  For $\ficm$ values below this critical fraction, we assume $v_e = 0$.

We expect the momentum conservation formulation to fail at low values of $\ficm$ (or low $v_{\rm wind}$) when we can't ignore the gravitational deceleration, but to be increasingly accurate for unbound gas.  On the other hand, when the energy formulation exceeds the momentum prediction, that implies that thermal pressure gradients are accelerating the gas and doing work above and beyond the momentum content of the inflowing gas. This seems unlikely when mixing is operating in the strong cooling limit, as high-resolution models of the hot/cold interface show no pressure gradient (Fielding et al. 2020). Therefore, we expect the energy conservation argument to fail for high values of $\ficm$. One simple way to combine this to estimates is simply to take the minimum predicted velocity of each:
\begin{equation}
\label{eq:vcloud}
v_{\rm cloud} = \min{(v_m, v_e)}
\end{equation}
It is this simple model that we will compare to simulations in the rest of this paper.

\begin{figure}
    \centering
    \includegraphics[scale=0.48]{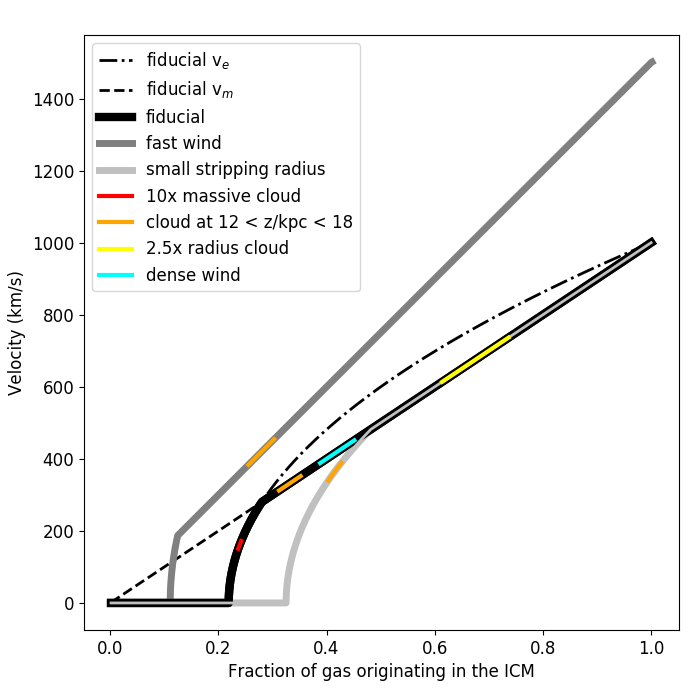}
    \caption{An illustration of our analytic model, showing the predicted gas velocity versus the fraction of gas originating in the ICM (a gas cloud at $\ficm=0$ only contains gas from the galaxy while one at $\ficm=1$ contains only ICM gas).  The thick solid line shows v$_{\rm cloud}(\ficm)$ from Equation~\ref{eq:vcloud}, while the dashed and dash-dot lines show where $v_m$ and $v_e$ are larger than $v_{\rm cloud}$, for the fiducial model (see text).  In addition to the fiducial model, we show a model with a fast wind (grey line) and one in which the gas is stripped from a smaller radius (silver line), where it must escape from a location deeper in the potential well.  We also use colored segments to show the locations along these tracks corresponding to the $\ficm$ value that gas at a height ranging from 12 to 18 kpc (above the disk) would have for different assumptions of the cloud mass and wind velocity, as noted in the legend. }
    \label{fig:cartoon}
\end{figure}

The cloud velocity in this model depends on the ICM fraction of the clouds ($\ficm$), in addition to the wind and escape velocities. We illustrate these relationships in Figure~\ref{fig:cartoon}.  The primary (thick black) curve in this plot shows the model just derived (Eq.~\ref{eq:vcloud}), with dashed and dot-dashed thin lines showing $v_m$ and $v_e$, respectively.  We also explore changing $\vwind$ and $\vesc$: a perusal of the three lines in this illustration shows that the velocity-ICM fraction relationship is affected in different ways by these parameters.  We vary the escape velocity by changing the cylindrical radius from which gas is stripped (called the stripping radius throughout the paper) --in the fiducial model (black) this is 20 kpc, and we reduce this to 2.5 kpc in the silver comparison line in the cartoon (a perusal of the galaxy potential described in Section \ref{sec:sim} and Tonnesen \& Bryan (2009) connects the disk radius to the escape velocity).  The wind velocity increases from 1000 km/s in the fiducial case to 1500 km/s for the ``fast wind" comparison (grey line).  We see that changing the radius from which gas is being stripped more strongly affects the velocity-ICM fraction relationship at low velocities and levels of mixing, as the silver line lies along the black line (seen as the narrower line).  However, changing the wind velocity increases the velocity of stripped gas at all mixed fractions.  

\begin{figure*}
    \centering
    
    \includegraphics[scale=0.77, trim=28mm 45mm 5mm 32mm,clip]{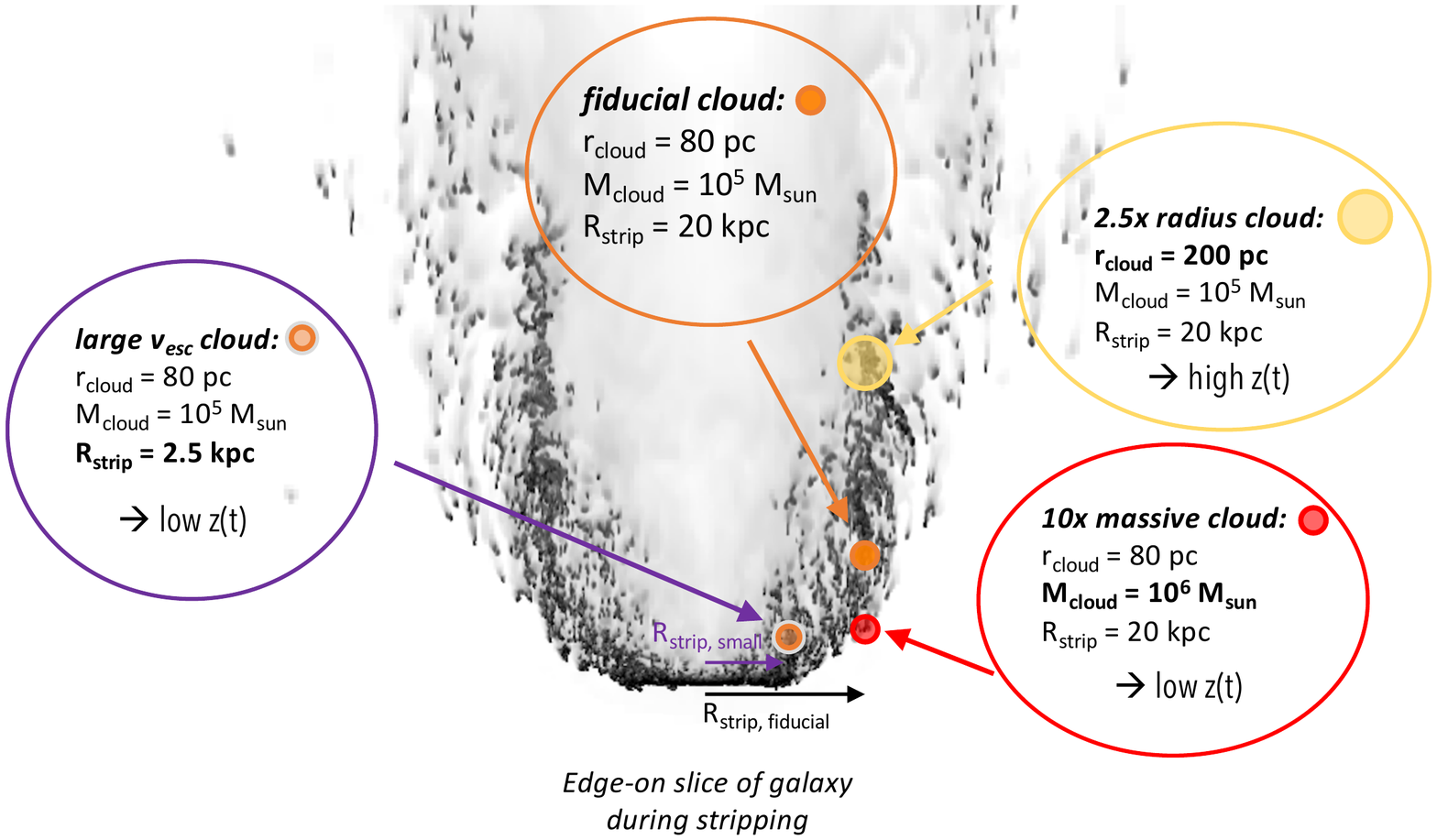}
    \caption{An illustrative cartoon showing the importance of cloud properties for the acceleration of gas from a galaxy due to ram pressure stripping, as predicted by Eq.~\ref{eqn:zt}.  For context, the background greyscale is the maximum density within a 2.4 kpc slab from run HDHV, taken at a single snapshot in time.   Cartoon ``clouds" (shown as colored circles) with different properties are depicted on top of this map and are labelled by their properties.  Each cloud differs from the fiducial cloud (orange) in one parameter, with the others held constant: in particular, we fix $\ficm$, and vary $v_{\rm esc}$ (orange and silver), mass (red), and radius (yellow); we have written how each variable affects the distance as a function of time ($z(t)$ from Equation \ref{eqn:zt}) for fixed $\ficm$. The colors are chosen to match those in Figure~\ref{fig:cartoon} (we use purple to differentiate the large $v_{\rm esc}$ parameters, although the cloud is orange and silver as in Figure~\ref{fig:cartoon}). Note that we are using this image {\rm only} as a background to show the context; the individual colored circles do not refer to specific clouds in this simulation.}
    \label{fig:illustration}
\end{figure*}

In its most simple form, as described above, our model does not predict the distance of the cloud from the disk, largely because this depends of the rate of mixing of the gas in the past.  However, if we assume a mixing rate for clouds we can make a straightforward prediction for the height of the cloud from the disk.  We make the simplest possible assumption, that clouds mix at a constant linear rate:
\begin{equation}
    \ficm(t) = \left(\frac{t}{t_{\rm final}}\right) \ficm(t_{\rm final})
\end{equation}
Here $t$ is the current time and $t_{\rm final}$ is the time at which the cloud reaches $\ficm(t_{\rm final})$.  Mixing occurs by the accretion of all of the wind that hits the cloud.  The time it takes to reach a chosen $\ficm(t_{\rm final})$ depends on the cloud mass ($M_{\rm cloud,final}$) and (effective) cross-sectional area ($A_{\rm cloud}$), as well as the wind velocity ($v_{\rm wind}$):
\begin{equation}
    t_{\rm final} = \frac{\ficm(t_{\rm final}) M_{\rm cloud,final}}{ \rho_{\rm wind} A_{\rm cloud} v_{\rm wind}}
\end{equation}

We note that our assumptions that all ICM mass that hits the cloud is accreted and that $A_{cloud}$ is constant with time are simplifications.  As discussed in Fielding et al. (2020) shear is likely important for mass transfer, as well as the cloud shape (Ji et al. 2019; Gronke \& Oh 2020).

Then, because we have the velocity as a function of the mixed fraction, we have the velocity of the cloud material\footnote{We stress that we are not trying to model the cloud as a single entity during its evolution, since the gas which ends up in a cloud at a given $t_{\rm final}$ may have a complicated previous history and is likely to come from multiple paths; therefore, these quantities should be thought as ``effective" values, characteristic of the cloud's gas history.} as a function of time.  
\begin{equation}\label{eqn:vel_tracerft}
v_{\rm cloud}(t) = \left\{ 
   \begin{array}{ll}
   \vwind \ficm \frac{t}{\tfinal} & t > t_2  \\
   \left[ \ficm \frac{t}{\tfinal} (\vwind^2 + \vesc^2) - \vesc^2 \right]^{1/2} & t_2 > t > t_1 \\
   0 & t_1 > t
   \end{array} \right.
\end{equation}
where $t_1 = \tfinal \vesc^2 / (\ficm (\vwind^2 + \vesc^2))$ is the time required for $\ficm(t)$ to reach the critical value mentioned earlier  ($f_{ICM,{\rm crit}}$), and $t_2 = \tfinal \vesc^2 / (\ficm \vwind^2)$ is the time when the momentum and energy predictions are the same (i.e. $v_e(t) = v_m(t)$).

Finally, we can solve for the distance ($z_{\rm cloud}(t)$) by integrating the velocity over time:
\begin{equation}\label{eqn:zt}
\begin{split}
z_{\rm cloud}(t) = \frac{2}{3} 
(\frac{\vesc^2\ficm}{q})^{1/2} \tfinal \left[ \left(\frac{t_2}{\tfinal} - \frac{q}{\ficm}\right)^{3/2}\right] \\ 
    + \frac{\vwind \ficm}{2 \tfinal} \left(t^2 - t_2^2\right)\\
\end{split}
\end{equation}
where $q = \vesc^2/ (\vwind^2 + \vesc^2)$.  If $t < t_2$, then the last term is dropped and $t_2 \rightarrow t$.

These relationships are illustrated in Figure~\ref{fig:illustration}, where we show the impact on the position of a fiducial cloud as we vary one property of the cloud at a time.  Unlike Figure~\ref{fig:cartoon}, this is a single snapshot, so the height of each cloud varies depending on its properties.  When an ICM wind has been acting on a galaxy, gas at a smaller radius is deeper in the galaxy potential and therefore has higher $\vesc$ (as discussed in Section \ref{sec:sim} and Tonnesen \& Bryan 2009), so in the same amount of time (and for fixed $\ficm$) it will not make it as far as our fiducial cloud.  In the paper we call this the stripping radius, and in Figure~\ref{fig:illustration} this is shown as R$_{\rm strip}$.  Also, a more massive cloud will be closer to the disk (red in Figure \ref{fig:illustration}), while a larger radius cloud (with fixed mass) will be farther from the disk (yellow in Figure \ref{fig:illustration}).

Using these simple assumptions, we can return to Figure~\ref{fig:cartoon} and determine where along the ICM-fraction - velocity track clouds will be as a function of height, using equation~\ref{eqn:zt}. In our schematic diagram we illustrate this relation using the height range from 15 kpc.  We vary the three parameters as listed in Figure~\ref{fig:illustration} so as to show the impact of $\vesc$, cloud radius, and cloud mass, modifying one parameter at a time while keeping the others fixed.  The orange bars describe 10$^5$ M$_{\odot}$ clouds with 80 pc radii being accelerated by the fiducial density wind.  Less massive clouds mix and accelerate more quickly than heavier clouds (compare the orange to red regions along the black fiducial relation), and clouds with larger radii mix more quickly than smaller ones (compare the orange to yellow regions). 

In addition, in Figure~\ref{fig:cartoon} we change two wind parameters: the wind velocity and density.  We have already discussed that changing the wind velocity changes the whole velocity-ICM fraction track, and we also see that this results in gas moving more quickly at any height above the disk (compare the two orange regions in the black and grey lines).  Also, with a faster wind, less mixed-in ICM mass is required to accelerate the gas to the same height, so the ICM fraction of a cloud is reduced with respect to a cloud.  Increasing the density of the wind increases the rate of mixing and energy/momentum input into the cloud, so given the same wind velocity a denser wind will result in clouds moving more quickly and being more well-mixed at any height above the disk (compare the orange and cyan regions along the black fiducial cloud relation).

Having developed and explored this simple model, in the next section, we carry out a set of high-resolution simulations of ram-pressure stripping with three different wind parameters and explore how well it performs.

\section{Method}\label{sec:method}

To follow the gas, we employ the adaptive mesh refinement (AMR) code Enzo (Bryan et al. 2014) which solves the fluid equations including gravity and optically thin radiative cooling.  The code begins with a fixed set of static grids and automatically adds refined grids as required in order to resolve important features in the flow.  Our simulated region is 300 kpc on a side with a root grid resolution of 256 cells.  We allow an additional 5 levels of refinement, for a smallest cell size of 37 pc.  The refinement criteria is based on gas mass, with a resolution of $\sim 1.9 \times 10^4$ M$_{\odot}$ (HDHV), $\sim 2.9 \times 10^4$ M$_{\odot}$ (HDLV), and $\sim 3.5 \times 10^4$ M$_{\odot}$ (LDLV), meaning that whenever a cell exceeds this mass it is refined into 8 small sub-cells.  The simulation includes radiative cooling using the GRACKLE (Version 3) cooling tables including metal cooling and the UV background from HM2012 (Smith et al. 2017).

\subsection{Simulation Initialization}\label{sec:sim}

Our galaxy is placed at a position corresponding to (150,150,75) kpc from the corner of our cubical 300 kpc computational volume, so that we can follow the stripped gas for more than 200 kpc.  The galaxy remains stationary throughout the runs, with the disk aligned in the x-y plane. The ICM wind flows along the z-axis in the positive direction, with the lower x, y, and z boundaries set for inflow and upper x,y, and z boundaries set as outflow.

We model a massive spiral galaxy with a flat rotation curve of 200 km s$^{-1}$. It includes a gas disc that is resolved to the maximum level (37 pc).  The galaxy model also includes the static potentials of the stellar disc, stellar bulge and dark matter halo, directly following the set-up of Roediger \& Bruggen (2006). Specifically, we model the stellar disc using a Plummer-Kuzmin disc (see Miyamoto \& Nagai 1975), using a radial scale length of 3.5 kpc, a vertical scale length of 0.7 kpc and a total mass of $1.15 \times 10^{11}$ M$_{\odot}$.  The stellar bulge is modeled using a spherical Hernquist profile (Hernquist 1993) with a scale length of 0.6 kpc and a total mass of 10$^{10}$ M$_{\odot}$.  The dark matter halo is modeled using the spherical model of Burkert (1995), with an equation for the analytic potential as given in Mori \& Burkert (2000).  The dark matter halo has a scale radius of 23 kpc and a central density of $3.8 \times 10^{-25}$ g cm$^{-3}$. We describe our disk setup in detail in Tonnesen \& Bryan (2009, 2010). 

To identify gas that has been stripped from the galaxy we also follow a metallicity value that is initially set to 1.0 inside the galaxy and 0.3 outside.  Because we do not include star formation in this simulation, the metallicity can also be used to determine the origin of the gas in each cell, in particular we can track the ratio of galactic gas to ICM gas on a cell-by-cell basis.

In this paper we discuss three simulations.  In all three runs, as in our earlier work (Tonnesen \& Bryan 2009, 2010, 2012), we impose a delay of 100 Myr before the wind enters the box in order to allow multiphase gas to self-consistently develop in the disk via radiative cooling. All of the ICM winds have temperatures of 7.08 $\times$ 10$^7$ K.  We vary the velocity and density as shown in Table~\ref{tbl:winds}.  The low density, low velocity wind (LDLV) has a velocity of 1000 km s$^{-1}$ ($\mathcal{M}\sim$ 0.79) and a density of 5 $\times$ 10$^{-28}$ g cm$^{-3}$.  The high density, low velocity wind (HDLV) has a velocity of 1000 km s$^{-1}$ ($\mathcal{M}\sim$ 0.79) and a density of 1.2 $\times$ 10$^{-27}$ g cm$^{-3}$. Finally, the high density, high velocity wind (HDHV) has a velocity of 3230 km s$^{-1}$ ($\mathcal{M}\sim$ 2.5) and a density of 1.2 $\times$ 10$^{-27}$ g cm$^{-3}$.  The minimum and maximum ram pressure parameters were chosen to roughly correspond to the ICM wind parameters of two jellyfish galaxies observed in the GASP sample (GAs Stripping Phenomena in Galaxies with MUSE; Poggianti et al. 2016): JO204 (LDLV) and JO201 (HDHV) (Gullieuszik et al. 2017; Bellhouse et al. 2017), with the middle simulation a test of the impact of changing a single wind variable rather than both density and velocity.

\begin{table}
\begin{center}
\label{tbl:winds}
\begin{tabular}{ c|c|c } 
 Name & Velocity & Density \\
  & km s$^{-1}$ & g cm$^{-3}$\\
 \hline
 LDLV & 1000 & 5 $\times$ 10$^{-28}$\\
 HDLV & 1000 & 1.2 $\times$ 10$^{-27}$\\
 HDHV & 3230 & 1.2 $\times$ 10$^{-27}$\\

\end{tabular}
 
\caption{The wind velocity and density of the three simulations discussed in this paper. }
\end{center}
\end{table}

\subsection{Cloud Selection}\label{sec:selection}

\begin{figure}
\centering
\includegraphics[scale=0.5]{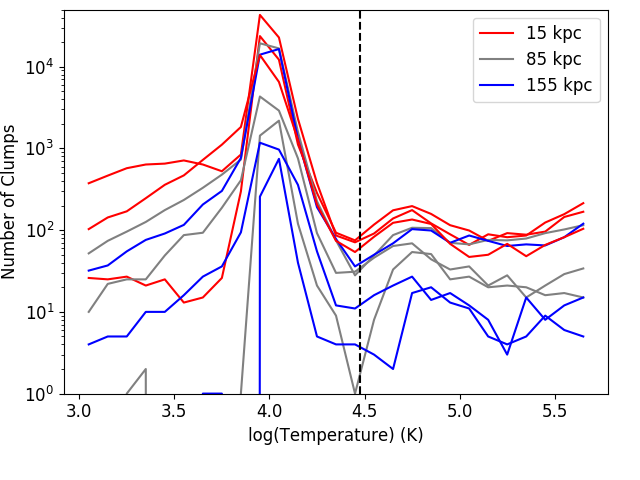}
\caption{The distribution of the minimum temperature found in clumps in the all three simulations (number of clouds per logarithmic temperature bin).  The colors denote height above the galaxy disk.  We have not differentiated between the three simulations as they all have a similar minimum in their temperature distribution. The y-axis scale is set to log to highlight the minimum in the distribution at all heights above the disk, at about 30 000 K (dashed vertical line). Clouds are required to have a minimum temperature below this value.}\label{fig:alltemps}
\end{figure}

In this paper, we examine both the state of all of the gas in the wake (Section~\ref{sec:allgascomparison}) as well as focusing just on the dense gas (Section~\ref{sec:dense}), which is more observationally accessible. 

In order to find higher-density ``clouds" in our tails, we used the clump finder routine in yt (Turk et al. 2011).  This uses level sets to find connected cells with values above a given selection criteria.  We searched for clouds using the gas density, starting with 10$^{-26}$ g cm$^{-3}$ and increasing the density by a factor of two for each level set.  This minimum density is relatively arbitrary, and we use other characteristics of the clumps, as described below, to make a more physically-motivated cloud selection.  In brief, we required a minimum number of cells and a low minimum temperature in the clump to include it in our analysis.

In more detail, the clump finder algorithm will find clumps down to a single cell, and indeed, we find that about half of our raw clump sample have fewer than 10 cells.  In order to eliminate clumps that are actually small density perturbations in the diffuse stripped tail, we only keep those identified with at least 10 cells in total (we note that we have repeated this work using clumps with at least 300 total cells and find the same trends, with some results shown in Appendix \ref{app:bigclouds}).  

We also find that clumps have a bimodal distribution in their minimum temperature, as shown in Figure~\ref{fig:alltemps}.  As we are attempting to choose cold, dense clouds rather than simple overdensities in the tail gas, we only include clumps whose minimum temperature is below 30 000 K.

We note that, when we impose these two selection criteria, the lowest maximum density of any cloud across all three simulations is more than 4 $\times$ 10$^{-26}$ g cm$^{-3}$, so our minimum search density (10$^{-26}$ g cm$^{-3}$) does not have a strong impact on the number of clouds we identify.  To orient the reader, a density of 4 $\times$ 10$^{-26}$ g cm$^{-3}$ will be refined to $\sim$300 pc, and only densities reaching $\sim$4 $\times$ 10$^{-23}$ g cm$^{-3}$ will be refined to 37 pc, i.e. the centers of the most dense clouds.

During the clump finding procedure we also save several cloud properties.  We find the maximum and minimum values of density, temperature, and the ICM fraction within the cloud.  Using all of the cells identified as belonging to our clouds, we also save the mass-weighted mean cloud position and velocity in addition to the physical characteristics listed above. 

In summary, our final set of ``clouds" are ten or more connected cells consisting of gas at least an order of magnitude denser than the ICM that have minimum temperatures more than three orders of magnitude lower than that of the ICM. This makes us confident that our clouds are physically distinct entities rather than just being ephemeral overdensities in the wake.

We identify these clouds in three evenly spaced regions:  12-18 kpc above the disk, 82-88 kpc above the disk, and 152-158 kpc above the disk.  Throughout the paper these heights will be identified as 15 kpc, 85 kpc, and 155 kpc.  These regions were selected so that we can easily compare gas properties as a function of height above the disk, and 
because they span the range of tail lengths seen in observations.  We tested our results using different height bins (10-20 kpc, 76-86 kpc, and 160-170 kpc) with no qualitative change to our results.

We perform this clump identification at each output (10 Myr apart) for each simulation.  We do not attempt to follow any individual clump, but identify the population of clumps in our simulations in the three regions at each output.

\begin{figure*}
    \centering
    \includegraphics[scale=0.25,trim= 25.mm 33mm 61.8mm 5mm, clip]{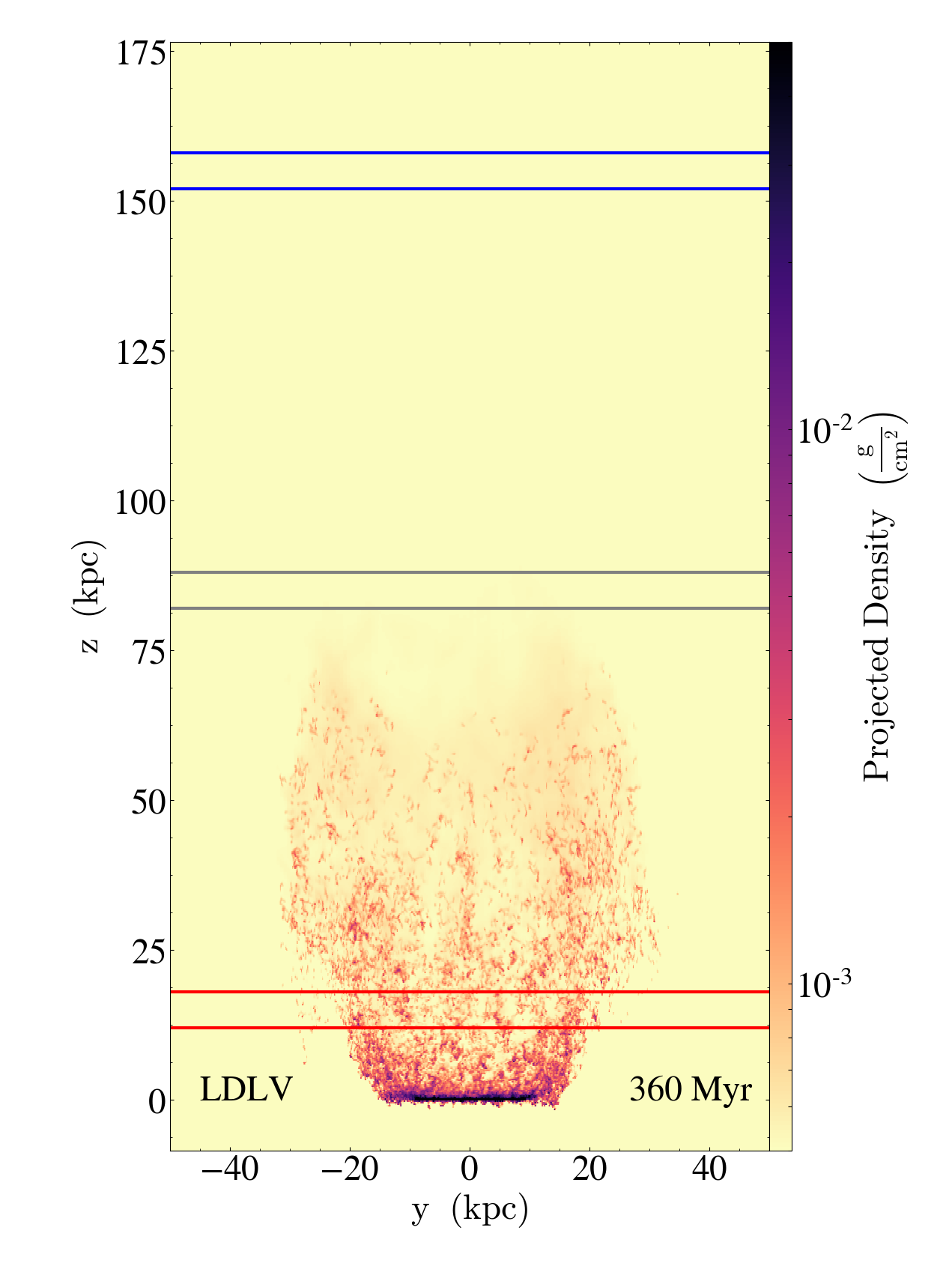}
    \includegraphics[scale=0.25,trim= 25.mm 33mm 61.8mm 5mm, clip]{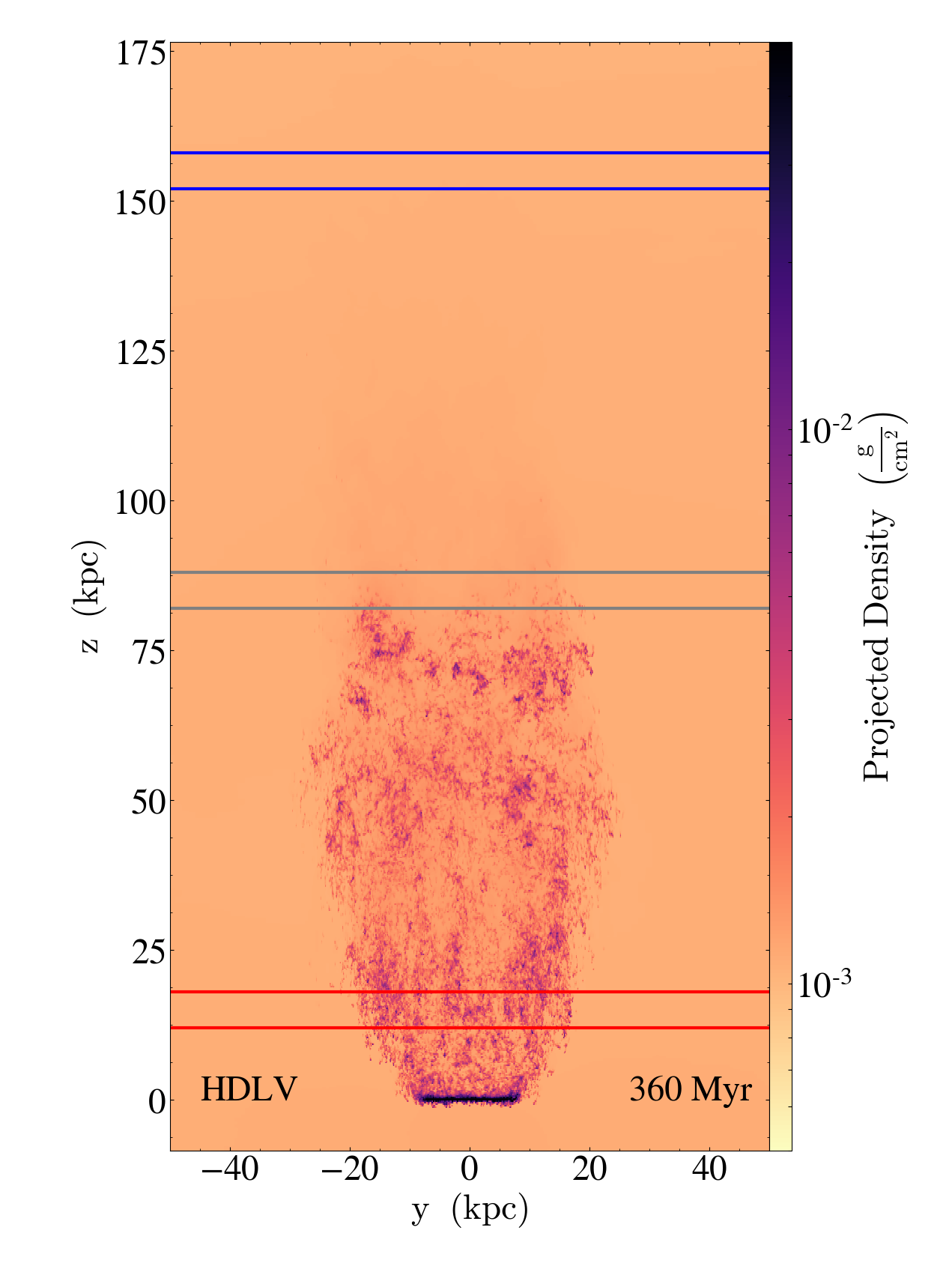}
    \includegraphics[scale=0.25,trim= 25.mm 33mm 61.8mm 5mm, clip]{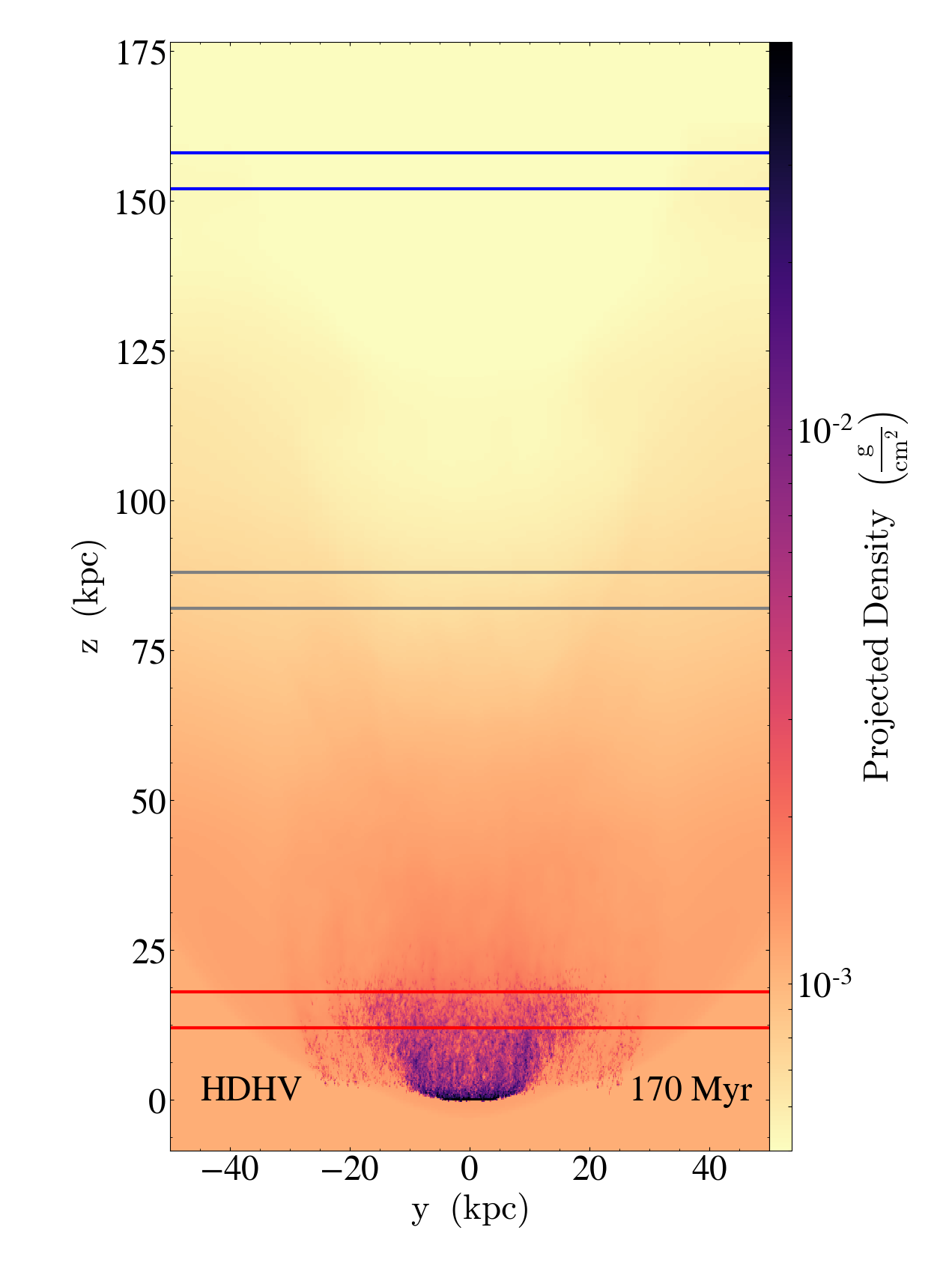}\\    
    \includegraphics[scale=0.25,trim= 25.mm 18mm 61.8mm 5mm, clip]{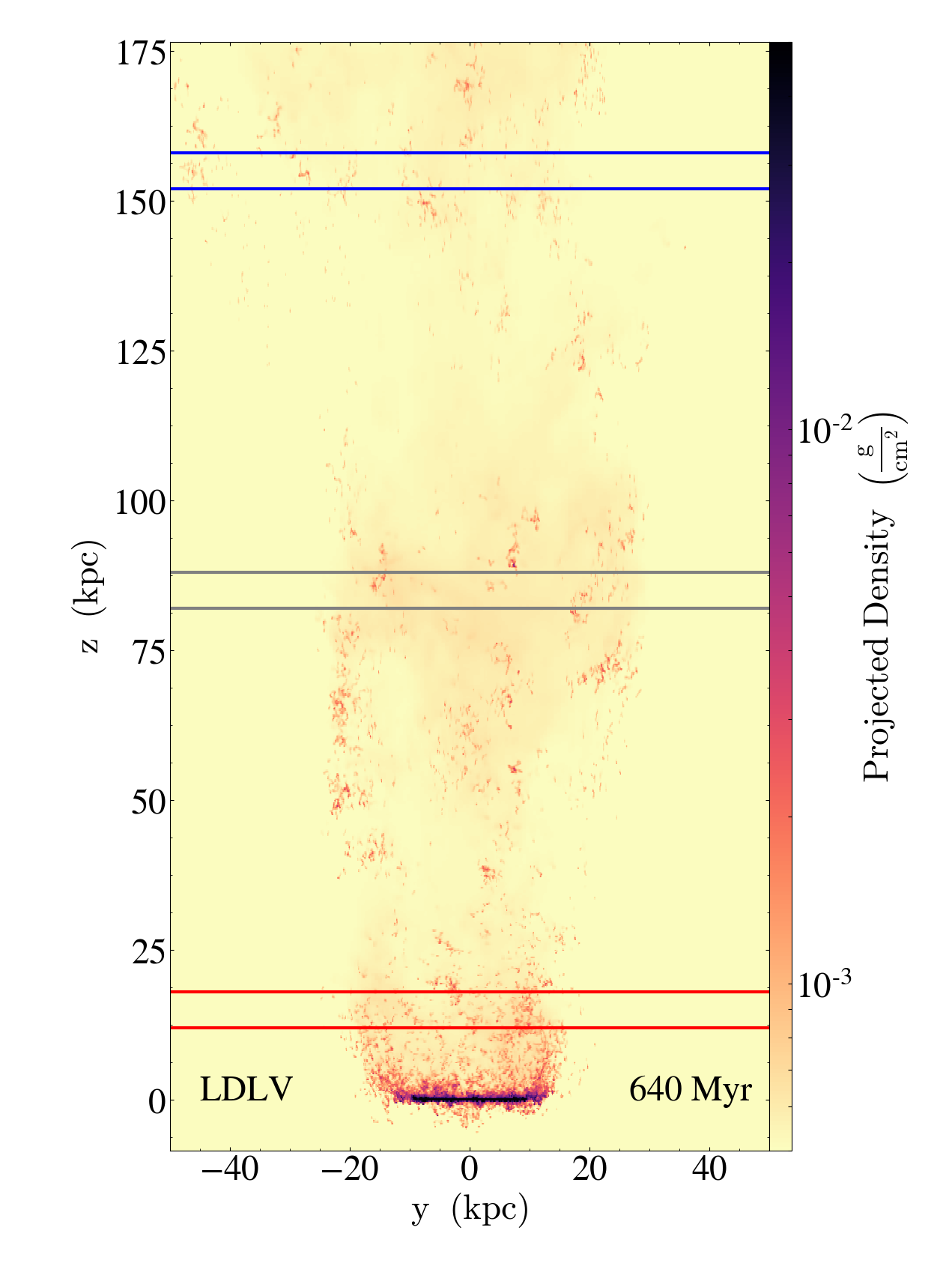}
    \includegraphics[scale=0.25,trim= 25.mm 18mm 61.8mm 5mm,clip]{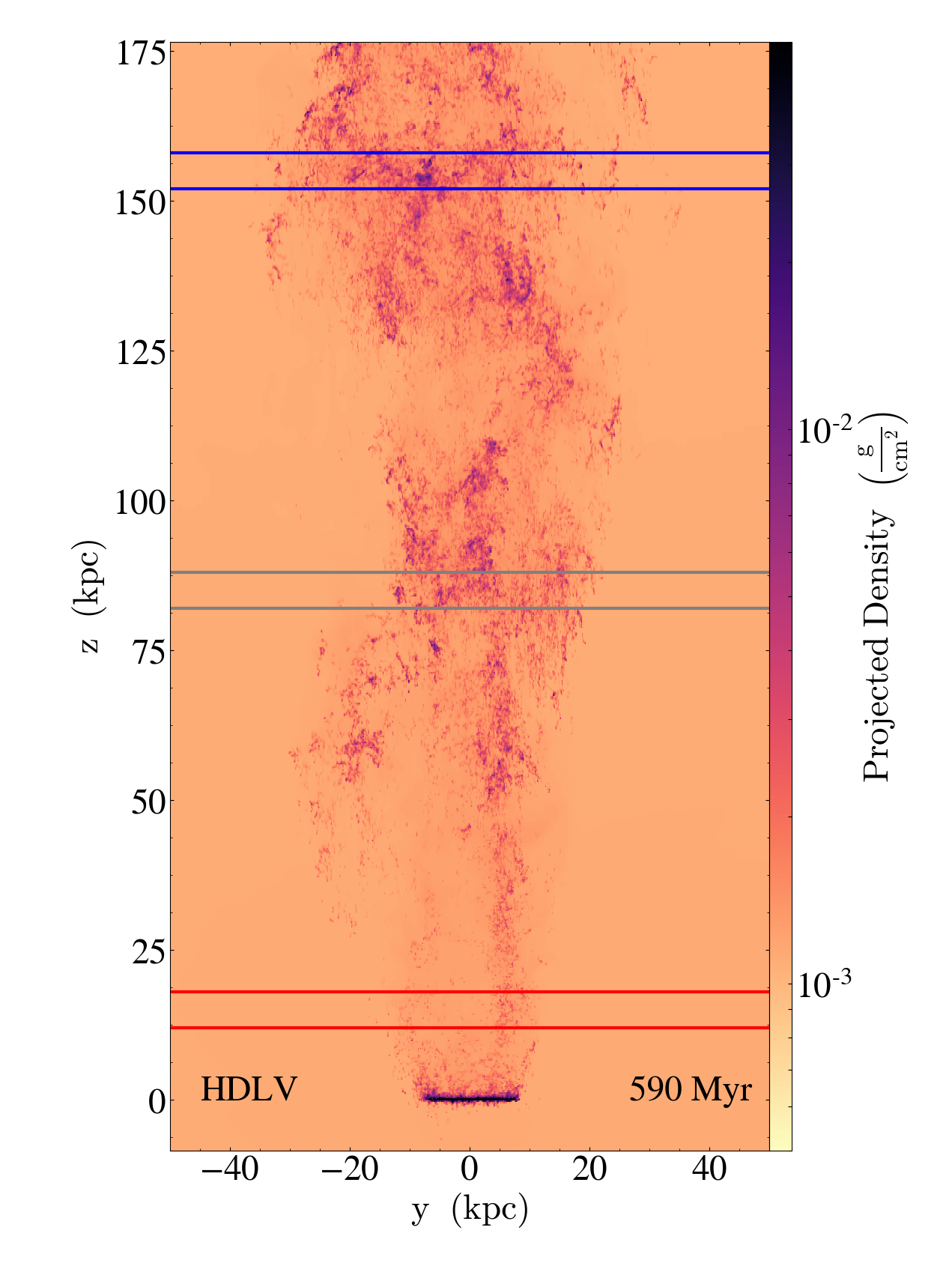}
    \includegraphics[scale=0.25,trim= 25.mm 18mm 61.8mm 5mm, clip]{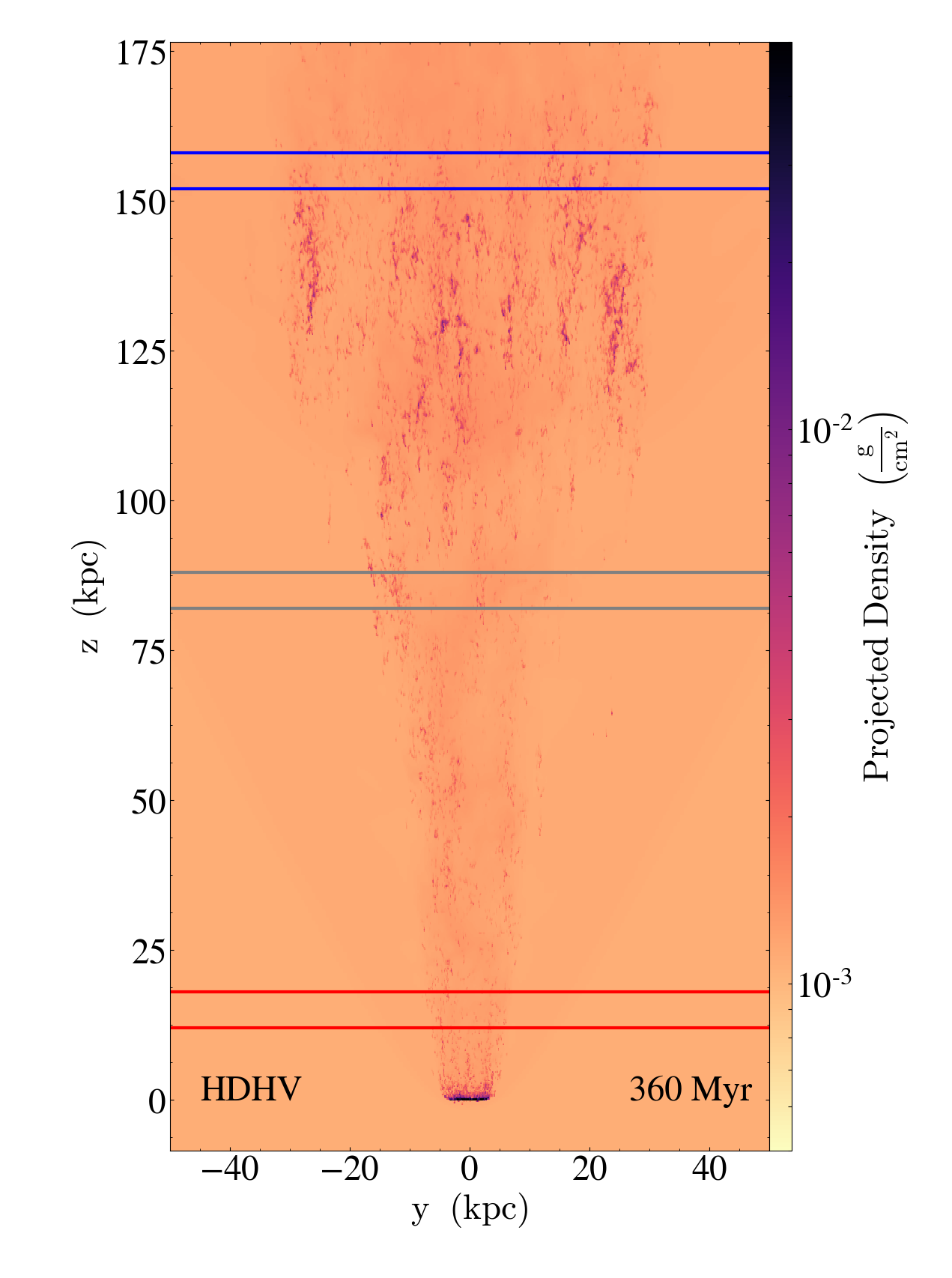}\\
    \includegraphics[scale=0.22,angle=270,trim= 260.5mm 5mm 12mm 5mm, clip]{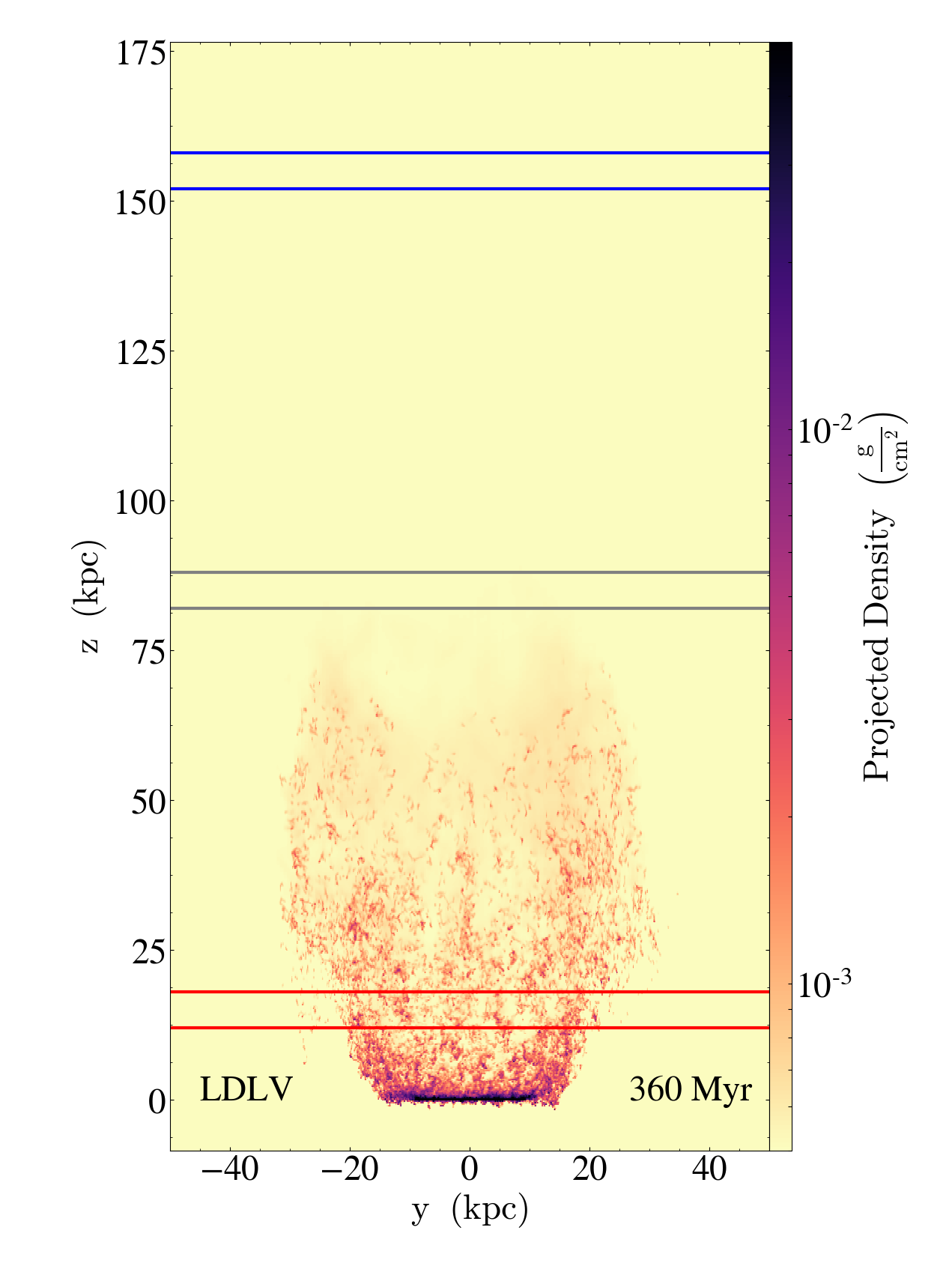}\\
    \caption{Images of projected density for an early (upper panels) and late (lower panels) output from each of our simulations.  Each projection is 185 kpc by 110 kpc.  The regions on which we focus in this paper are denoted by the colored lines (which are reflected in later figures).  Gas behind galaxies shows a range of densities and tail morphologies as a function of height above the disk, time since stripping began, and wind paramaters.}
    \label{fig:projections}
\end{figure*}

In Figure~\ref{fig:projections} we show density projections of ``early" and ``late" illustrative outputs from each simulation, with colored lines indicating the three narrow height ranges we analyze, as described above.  Here we are using ``early" and ``late" to denote the timeline of the development of the tail.  Specifically, the ``early" output is close to the output at which the most clouds are found in the 15 kpc region, and the ``late" output is close to the output at which the most clouds are found in the 155 kpc region (as seen in Figure~\ref{fig:numclouds}).  In order to directly compare the simulations, and highlight differences in the tails, we chose to show the 360 Myr snapshot for all three.  This same output is also used in Figures \ref{fig:ICM_vz_height} - \ref{fig:ICM_vz_runs}. 

The tail structure varies as a function of height and time within a simulation, and differs across simulations at the same time. At earlier times clouds are tightly packed in the tails, while at later times clouds are less densely distributed.  This is true even looking at a single height above the disk (for example within the red 15 kpc region).  The wind properties have a significant impact: for example, at 360 Myr, dense gas in the HDHV tail has reached 155 kpc from the disk, while in the LV runs it has barely reached 85 kpc.  Even within the LV runs the gas distribution is different:  denser clumps are seen at larger distances in HDLV than in LDLV.  We also note that the ICM of the HD runs is denser than in LDLV, leading to the darker background density seen in the projections.

\section{Testing the Mixing Model}\label{sec:comparison}

Our simple analytic mixing model, developed in Section~\ref{sec:mixing}, predicts a relationship between stripped cloud velocity and mixed fraction, modulo wind velocity and galaxy escape velocity.  In this section we use the three simulations described in Section~\ref{sec:sim} to test these predictions.

\subsection{All Gas}\label{sec:allgascomparison}

We begin by looking at the properties of all gas in the tail before turning to dense clouds (which connect more directly with observations) in the next section.
The gas velocity in our mixing model depends only on the ICM fraction of gas (for fixed $\vesc$ and $\vwind$), and therefore can be tested using all of the gas behind the galaxy in our simulations.  According to our model, this ``tail" gas will consist of a range of ICM fractions, from nearly pure galactic gas that was loosely bound to the galaxy and moving slowly, all the way to nearly pure ICM gas moving at the wind speed.  

\begin{figure}
    \centering
    \includegraphics[scale=0.64,trim= 3mm 12.5mm 39mm 13mm, clip]{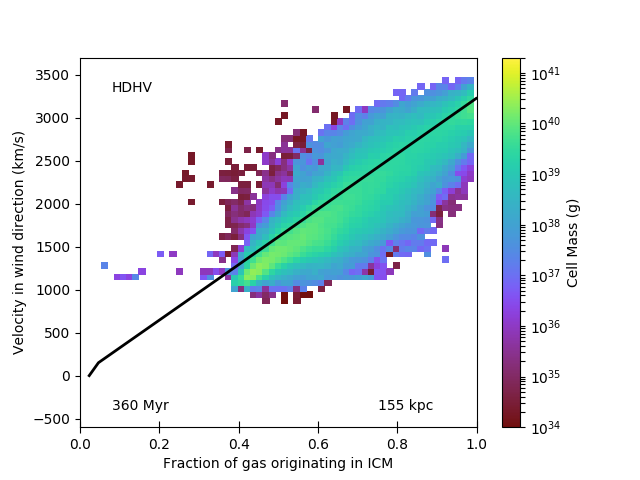}
    \includegraphics[scale=0.64,trim= 3mm 12.5mm 39mm 14.5mm, clip]{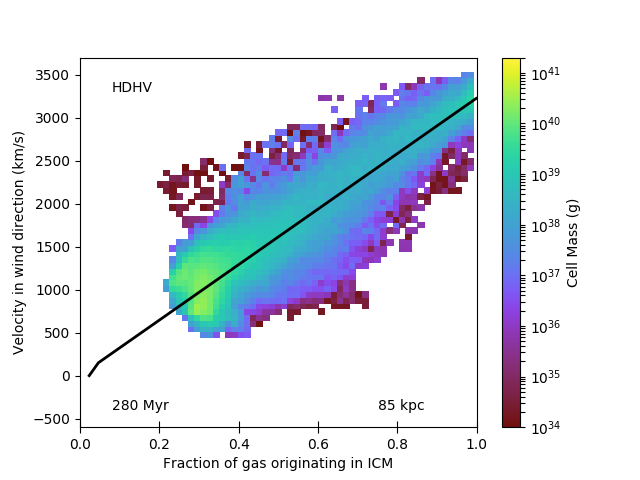}
    \includegraphics[scale=0.64,trim= 3mm 2.4mm 39mm 14.5mm, clip]{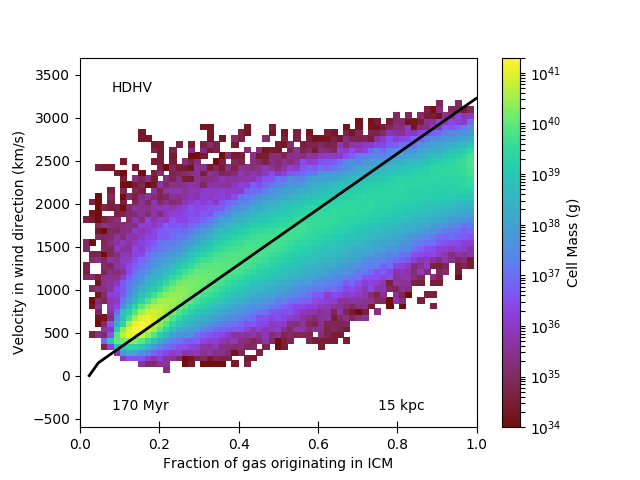}
    \includegraphics[scale=0.7,angle=270, trim= 127.5mm 1mm 13mm 13mm, clip]{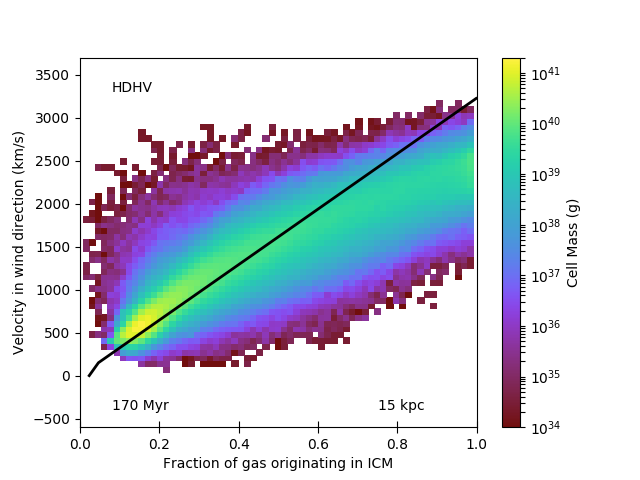}
    \caption{The gas velocity-$\ficm$ relation at three different heights (and times), as labelled, above the disk in run HDHV.  
    There is more mixing of gas as we move farther from the disk.  Despite this, the gas at all heights falls along a similar velocity - ICM fraction line. The solid line in each panel is from equation~\ref{eq:vcloud}.}
    \label{fig:ICM_vz_height}
\end{figure}

\begin{figure}
    \centering
    \includegraphics[scale=0.64,trim= 3mm 12.5mm 39mm 13mm, clip]{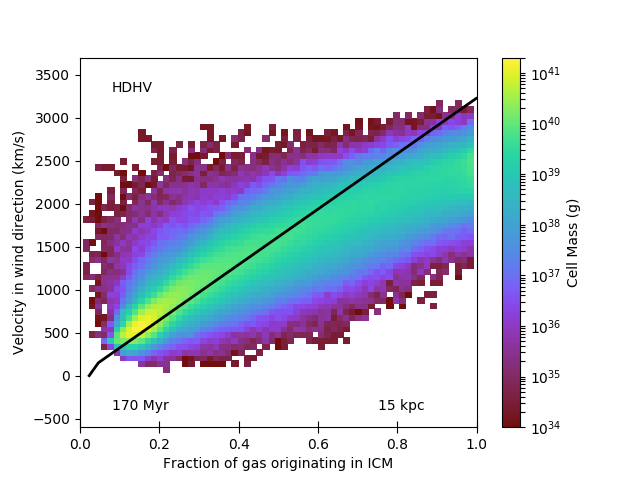}
    \includegraphics[scale=0.64,trim= 3mm 2.4mm 39mm 14.5mm, clip]{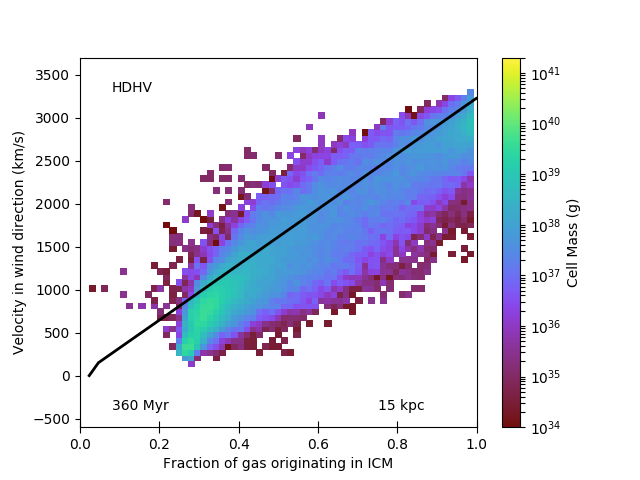} \\
    \includegraphics[scale=0.7,angle=270, trim= 127.5mm 1mm 13mm 13mm, clip]{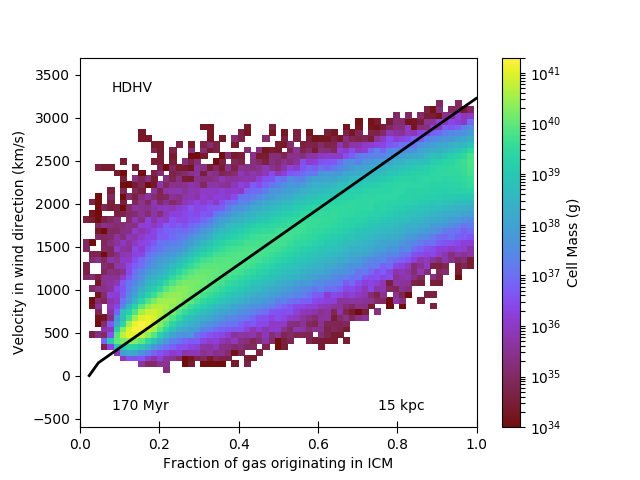}
    \caption{All gas between 15 kpc in HDHV at an early time (170 Myr) and later time (360 Myr).  The black line shows the model prediction using a stripping radius of 3 kpc.  At early times there is gas that is less mixed than at later times.  More energy is required to remove gas from inner radii, which results in the higher ICM fractions at later times.}
    \label{fig:ICM_vz_time}
\end{figure}

\begin{figure*}
    \centering
    \includegraphics[scale=0.64,trim= 3mm 12.5mm 39mm 13mm, clip]{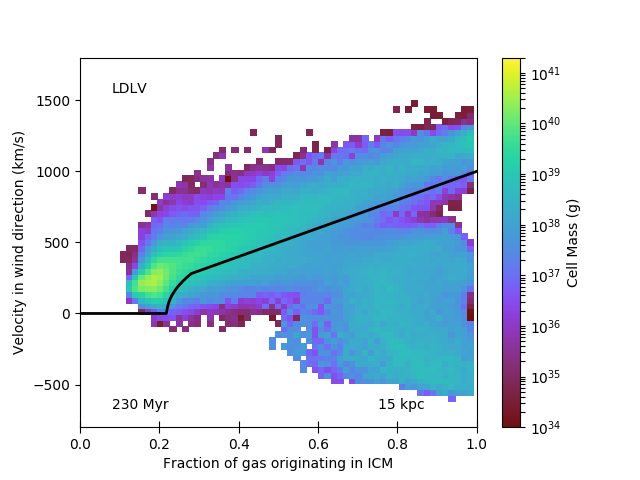}
    \includegraphics[scale=0.64,trim= 3mm 12.5mm 39mm 14.5mm, clip]{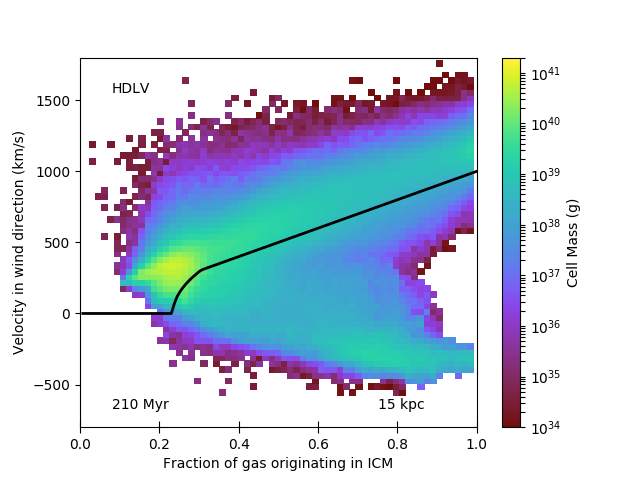} \\
    \includegraphics[scale=0.64,trim= 3mm 2.4mm 39mm 14.5mm, clip]{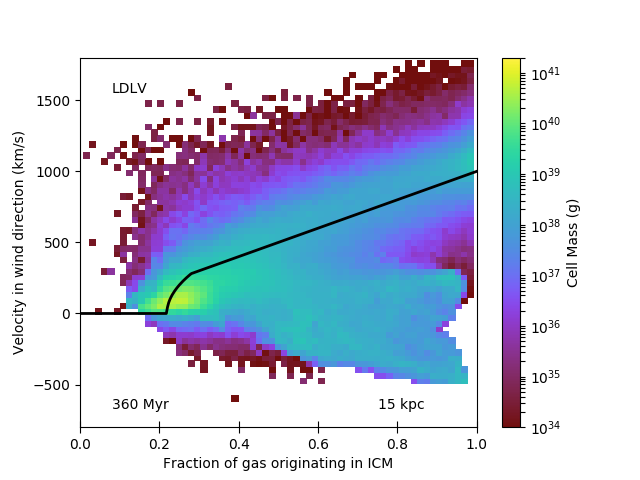}
    \includegraphics[scale=0.64,trim= 3mm 2.4mm 39mm 14.5mm, clip]{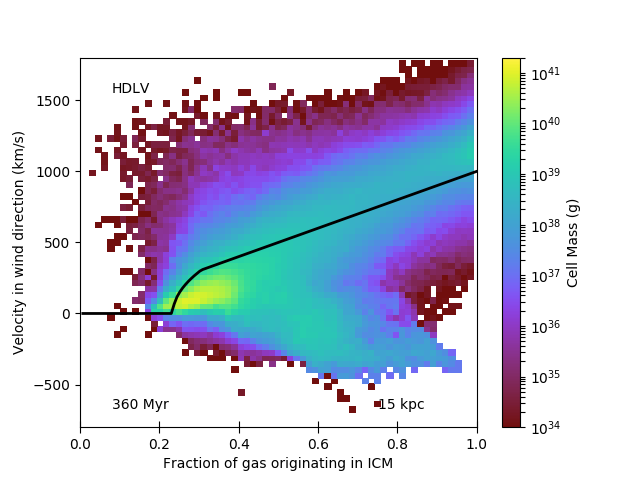}
    \includegraphics[scale=0.7,angle=270, trim= 127.5mm 1mm 13mm 13mm, clip]{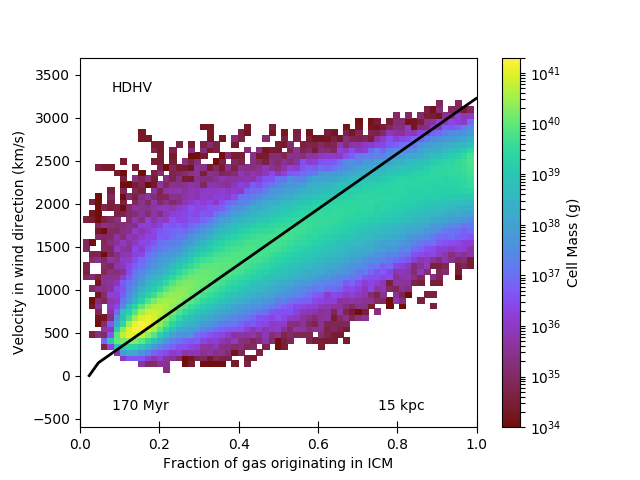}
    \caption{Gas from 15 kpc at an early (top panels) and later (bottom panels) time in the LV runs: LDLV (230 Myr and 360 Myr) and HDLV (210 Myr and 360 Myr).  In each run the majority of gas falls near the analytic line.  Gas with low ICM fractions tends to be moving with higher velocities at earlier times.  Also, in comparison to the HDHV run, the LV gas tails show a much broader velocity distribution at a given ICM fraction.}
    \label{fig:ICM_vz_runs}
\end{figure*}

In Figure~\ref{fig:ICM_vz_height}, we see that, indeed, gas behind the galaxy has a range of ICM fractions.  In this figure we plot the velocity-ICM fraction relation in the HDHV run at three heights above the disk (15 kpc, 85 kpc, and 155 kpc).  Each of these plots use an output taken shortly after any dense clouds are identified at those heights.  In black we overplot the analytic relation between ICM fraction and velocity using a stripping radius of 3 kpc (used to determine $\vesc$), as this is about the radius of the gas disk at the end of the simulation.  We see that at any height above the disk, gas tends to have a broad range of ICM fractions, and generally falls along the analytic relationship between ICM fraction and velocity. As we move farther from the disk, the minimum velocity and ICM fractions are shifted to higher values.  This agrees well with our model prediction that continual mixing occurs as gas is driven away from the disk, therefore our lowest ICM fractions will be found near the disk.

We note that the tail gas from 15 kpc in HDHV shows a maximum velocity ($\sim 2500$ km/s) that is lower than our input wind velocity (3230 km/s).  This is because the wind velocity in HDHV is supersonic and creates a large bow shock, deflecting the incoming wind along the x- and y-axes in addition to the z-axis.  At later times, much of the gas is stripped, the bow-shock shrinks, and gas flows more directly along the z-axis, the direction of gas inflow at the edge of the box.  As we show below, the slow velocities in the z-direction of high $\ficm$ gas is not a long-lived effect.

In Figure~\ref{fig:ICM_vz_time} we examine how the velocity-ICM fraction relation changes over time in our simulations. We plot this relation in the HDHV run 15 kpc above the disk at an early (170 Myr) and late (360 Myr) time.  The amount of gas in the tail changes (as we will explore in more detail later in the paper), but here we highlight the fact that, at all times in our simulations, most of the gas in the tail falls approximately along our analytic relation.

By examining the minimum ICM fraction of gas from 15 kpc as a function of time, Figure \ref{fig:ICM_vz_time} illustrates the importance of the escape velocity in the velocity-ICM fraction relation.  We find that at early times some gas is moving slowly with low ICM fractions, while at late times there is only gas moving quickly with higher mixed fractions.  This reflects the decreasing stripping radius, and the resulting increase in escape velocity, as a function of time.  The impact of the stripping radius on the ICM fraction and velocity of gas can be seen in Figure \ref{fig:cartoon} by comparing the orange bars in the fiducial and small stripping radius lines.

We note that at later times the velocity of low-$\ficm$ gas tends to fall below the analytic line.  In comparison with the larger distances in Figure \ref{fig:ICM_vz_height}, we find that this occurs most dramatically near the disk.  We posit two likely reasons for this lower velocity gas.  First, there continues to be a small bow shock near the disk, and the gas flow is not completely in the z-direction.  We also find that at early times the disk is being stripped rapidly and shrinking, but at late times the disk size stabilizes.  Therefore gas that is stripped is more likely to be near the disk edge and somewhat protected from the wind.  We discuss this more below. 

Now that we have carefully compared the HDHV run to our analytic model, in Figure~\ref{fig:ICM_vz_runs} we verify that this relationship holds in all three of our runs.  Here, in two columns we have plotted the gas 15 kpc above the disk in LDLV and HDLV, in each case choosing a time shortly after dense clouds are identified (upper panels) and at 360 Myr (lower panels) to match the projections in Figure \ref{fig:projections}.  The disk stripping radii for the model lines are chosen to be 20 kpc (LDLV) and 15 kpc (HDLV), as these are near the surviving disk radius at early times for the three runs.  We note that at early times, gas with a low ICM fraction is moving more quickly than the analytic line, likely due to gas stripped from a large galactic radii, while at 360 Myr more of the gas lies along or below the line, indicating stripping from small radii.  This agrees well with our model prediction illustrated in Figure \ref{fig:cartoon}.

The generally good agreement between the model predictions and the gas distribution in these plots demonstrate the basic success of the picture.  While the overall predictions are verified, we highlight two points of disagreement between the LV runs and our model predictions.  First, at the early times shown in Figure~\ref{fig:ICM_vz_runs}, the gas velocity is slightly larger than predicted, and this is true even for pure ICM gas (at the right edge of the upper two panels), which have speeds in excess of the inflow velocity (1000 km/s for the LV runs).  We see in the lower panels of Figure \ref{fig:ICM_vz_runs} that this excess velocity is less pronounced at later times.  The second, more dramatic, disagreement between the simulations and the model in the LV runs is the gas component with high ICM fractions and low, even negative, velocities (in the lower-right of each panel). 

\begin{figure}
    \centering
    \includegraphics[scale=0.315]{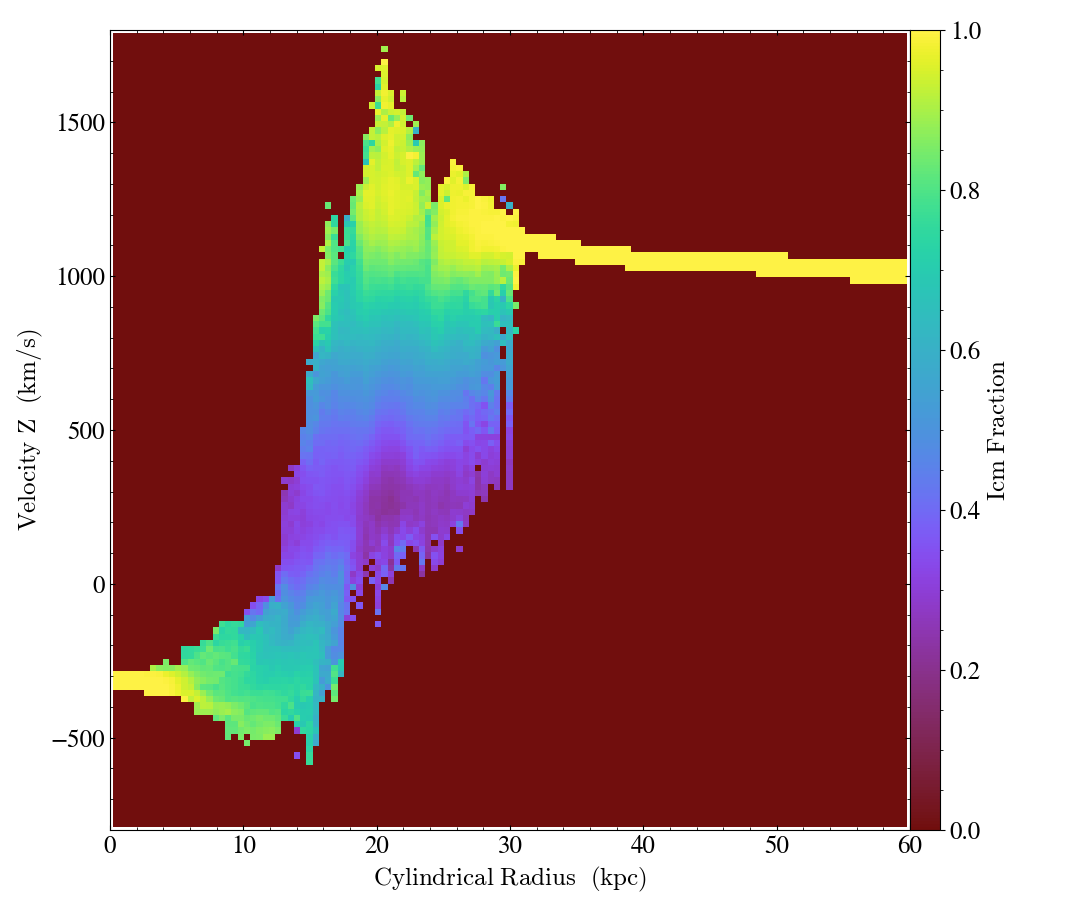}\\
    \includegraphics[scale=0.315]{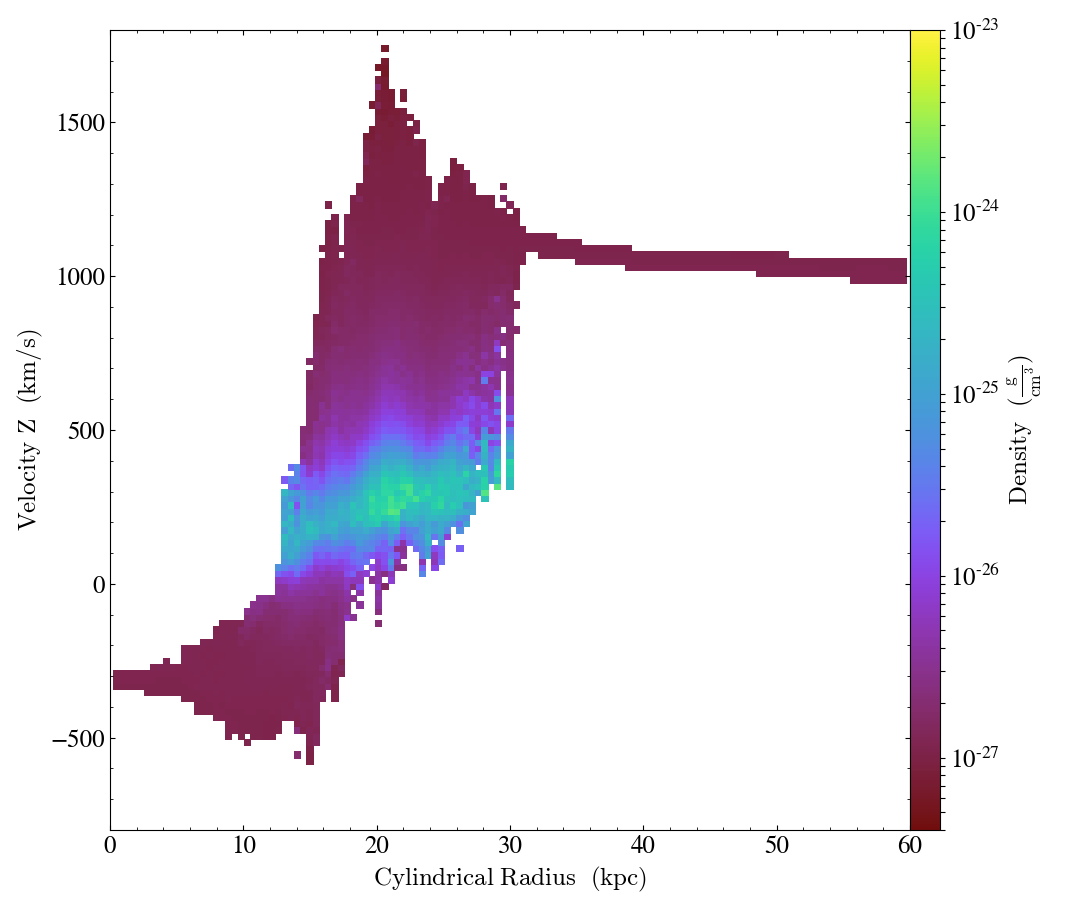}
    \caption{These panels both plot the gas velocity in the wind direction against the cylindrical radius (for all gas in the 15 kpc bin) from the disk in the HDLV simulation at 210 Myr into the simulation.  The upper and lower panels are color coded by the mean ICM fraction and gas density in each histogram cell, respectively.  We see that the high ICM fraction, low density gas in the inner tail region is moving quickly, and that the low density gas in the disk ``shadow" has negative velocities.}
    \label{fig:radius_density_comp}
\end{figure}

To look at these discrepancies in more detail, in Figure~\ref{fig:radius_density_comp} we plot the z-velocity of the gas, as a function of cylindrical radius for gas within the same 15 kpc height range of the disk in the HDLV run (to match the HDLV panel in Figure~\ref{fig:ICM_vz_runs}).  Rather than the total mass in each histogram cell, we have color-coded these two plots using the mean ICM fraction (top) and the mean gas density (bottom) in each cell of the histogram.  

First we describe the general gas distribution in these panels.  In the inner regions, gas only has negative velocities -- this indicates the disk's ``shadow", where it blocks the flow of the ICM wind.  There is then a region between $\sim$15-30 kpc with gas at a wide range of velocities, indicating that the majority of the tail is found at these radii, and finally at very large radii we see pure ICM gas at the inflow velocity.  

We now see that gas moving at the highest velocities (even higher than the input wind velocity) is near the border between the disk ``shadow" and the tail region, at the edge of the disk.  Visual inspection suggests that this is due to vortices created by the wind flowing past the surviving disk, with some smaller vortices created by dense clouds in the tail.  When the vortex velocity aligns with the wind velocity we see our maximal velocities.  At early times a particularly large vortex forms that results in a large amount of gas at high velocity while at later times the majority of ICM-like gas is moving at the inflow velocity. We do not see these fast flows in the HDHV run because the galaxy is stripped quickly to a small radius so there are not large vortices containing a significant amount of mass. 

We can also clearly see in Figure~\ref{fig:radius_density_comp} that gas with high ICM fractions and negative velocities is falling back in the shadow of the disk (i.e. at small galactocentric radius).  It falls off of our predicted $\ficm$-velocity relation because the primary mode of acceleration of this gas is gravitational acceleration rather than mixing. The resulting infall velocity is of order 300 km/s, a value which is set by the acceleration of the galaxy and is unrelated to the wind velocity.   The remaining gas disk is very small in HDHV, and so there is no fallback in that run (as seen by the lack of negative velocities in Figure~\ref{fig:ICM_vz_height}).  Interestingly, the gas in the shadow of the disk is generally low density, which likely allows it to be more easily shifted to different radii due to disordered motion or vortices (there may be occasional fallback of clouds but this is a minor effect in these runs).  When we focus on just the dense gas in the bottom panel, we see a very clear relationship between radius and velocity, again illustrating the importance of the stripping radius (and thus escape velocity) on gas acceleration. 

In summary, our simple model is an excellent prediction of the gas velocity-ICM fraction relation, and the differences between the model and simulations can be explained by the hydrodynamical interaction of a wind hitting a disk.

\subsection{Focusing on Dense Clouds} \label{sec:dense}

While it is commonly accepted that low-density gas in ram pressure stripped tails mixes with the ICM, dense clouds are often assumed to consist of galactic gas that maintains its integrity as it is being accelerated by the wind.  Therefore, in this section we focus directly on comparing our model predictions to dense clouds in the tail.

\begin{figure}
\includegraphics[scale=0.55,trim=0mm 15.2mm 0mm 5mm,clip]{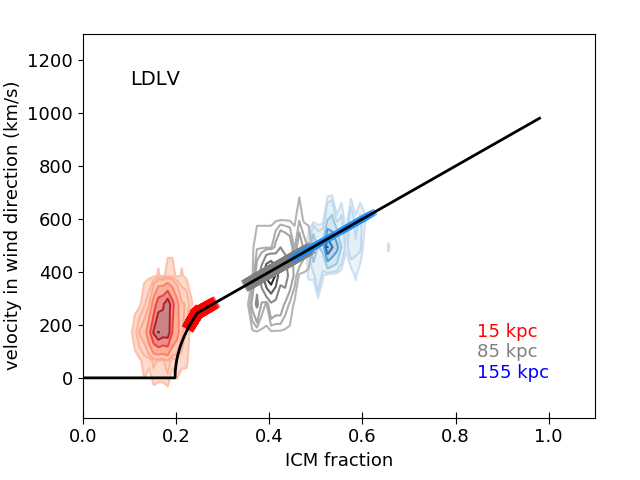}\\
\includegraphics[scale=0.55,trim=0mm 15.2mm 0mm 8.7mm,clip]{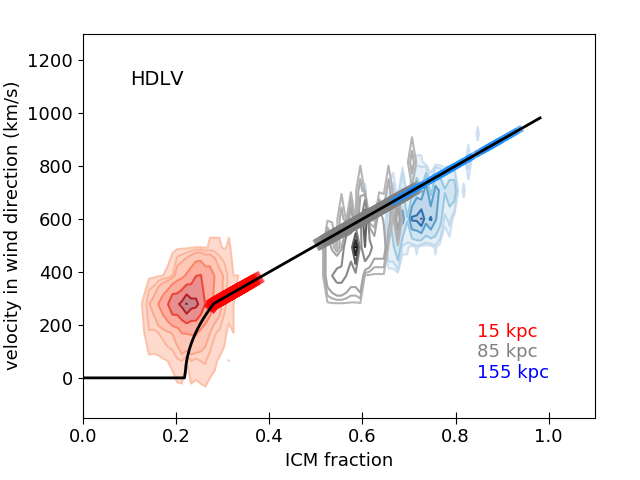}\\
\includegraphics[scale=0.55,trim=0mm 0mm 0mm 8.7mm,clip]{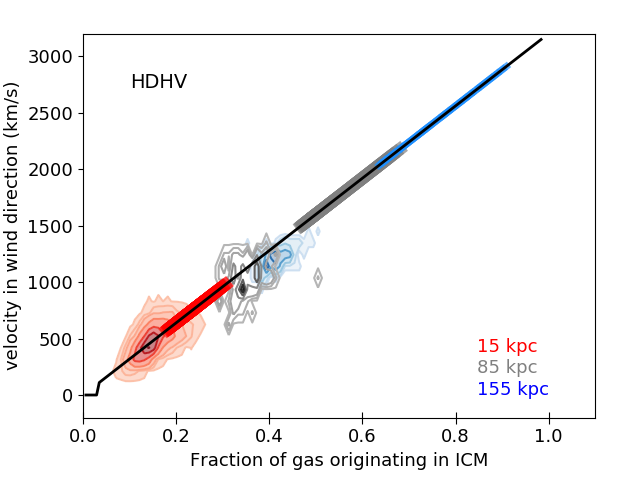}
\caption{The velocity as a function of the ICM fraction of clouds.  The contours show the density of clouds per linear range in velocity and gas fraction.  As in Figure \ref{fig:ICM_vz_runs}, the panels from top to bottom are the three runs: LDLV, HDLV, and HDHV. Clouds from the first 4 (15 and 85 kpc) or 5 (155 kpc) outputs at which they are identified in each simulation and height are shown.  For the analytic model, the cloud radius is set to 80 pc, and the mass ranges from 2$\times$10$^5$ to 4$\times$10$^5$ M$_{\odot}$.  For LDLV and HDLV the wind velocity is set to be 1000 km/s, and for HDHV the wind velocity is set to be 3230 km/s. Simulations are generally in good agreement with the analytic model.}

\label{fig:tracer_vz}
\end{figure}

As we did with all of the gas in our tail, we compare the cloud velocity to the ICM fraction in our clouds in Figure~\ref{fig:tracer_vz}.  The contours in each panel show the cloud velocity versus ICM fraction for clouds identified at early times in each simulation.  The clouds are defined using contiguous regions of dense gas, as described in Section \ref{sec:selection}, and the contours show the number of clouds in any bin.

We use clouds identified in a limited range of outputs because the radius from which gas is stripped decreases over time so the effective escape velocity for clouds at a given height increases over time.  As our model uses a single escape velocity, we choose a narrow range in time to decrease this variation.  We also note that we are not tracking individual clouds in our simulations, so we cannot determine the relationship between clouds at different distances and outputs.  Therefore, we use the first 4 outputs (5 at 155 kpc in order to have more than 150 clouds in LDLV) at each height that have any identified clouds in an attempt to follow a similar population of clouds moving away from the disk.

In order to overplot a model line we made several choices for parameters based on our simulations.  First, we chose a stripping radius in order to calculate the escape velocity.  We do this by computing the distance to the x$=$y$=$0 line (through the disk center). Based on this analysis, we identified stripping radii of 30 kpc, 20 kpc, and 10 kpc for LDLV, HDLV, and HDHV, respectively.  These values are meant to be representative of the radius from which gas has been stripped, and as such are smaller than the original disk radius but larger than the surviving disk radius.  Thus these values are smaller than the initial stripping radius and larger than the radius from which gas is stripped at late times.  Note that the line does not change dramatically due to the minor change in stripping radius between LDLV and HDLV.  The wind velocity and escape velocity sets the shape of our curves, as shown in Figure~\ref{fig:cartoon}. 

We also examined the size and mass of clouds and chose a cloud mass ranging from $2 \times 10^5$ to $4 \times 10^5$ M$_{\odot}$ for all three simulations.  This is based on the measured mass distribution of our selected clumps, which ranges from about $5 \times 10^4$ to $1 \times 10^6$ M$_{\odot}$, with the peak of the mass distributions in all simulations at about $2 \times 10^5$ M$_{\odot}$.  A cloud radius cannot be directly measured, as the clouds are not spherical.  Indeed, we expect them to be extended along the wind direction.  We choose a radius of 80 pc based on the measured volumes of the clouds, recognizing that this could vary by a factor of several (from 1 cell across to a radius of 600 pc, corresponding to a spherical cloud at our maximum measured volume of 3 $\times$ 10$^{64}$ cm$^3$).  Using these assumptions allows us to predict where along each curve the clouds will fall when 15 kpc, 85 kpc, and 155 kpc above the disk (Equations~\ref{eqn:vel_tracerft} \& \ref{eqn:zt}).  This is shown by the thick shaded regions along the curves.

Despite the simplicity of our model, we again find good agreement with the simulations. In particular, the $\ficm$-velocity relations follow quite well the predictions of the model and there is qualitative agreement in that higher cloud height corresponds to larger cloud velocities and higher mixed fractions.  However, there is some disagreement with the predictions in detail, which we can separate into two general trends: shifts above or below the model curve, and shifts along the model curve.
\begin{figure*}
\centering
\includegraphics[scale=0.57,trim= 0mm 17mm 0mm 1mm, clip]{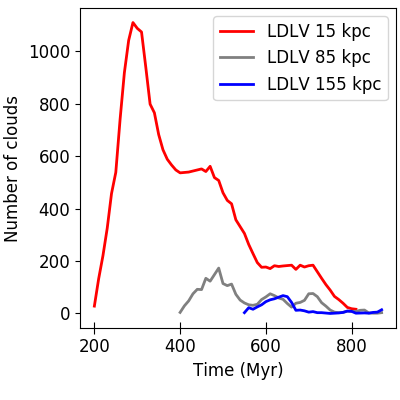}
\includegraphics[scale=0.57,trim= 0mm 17mm 0mm 1mm, clip]{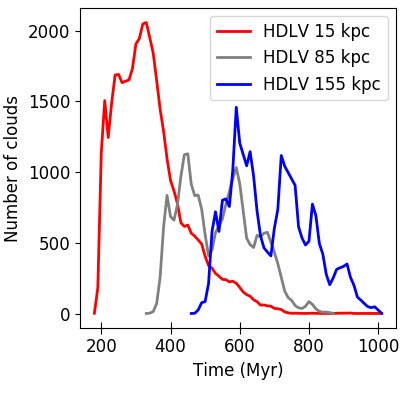}
\includegraphics[scale=0.57,trim= 0mm 17mm 0mm 1mm, clip]{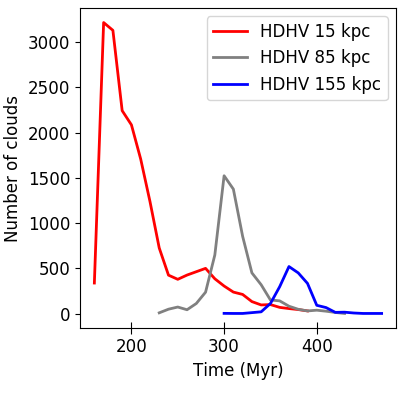}\\
\includegraphics[scale=0.57,trim= 0mm 2mm 0mm 2mm, clip]{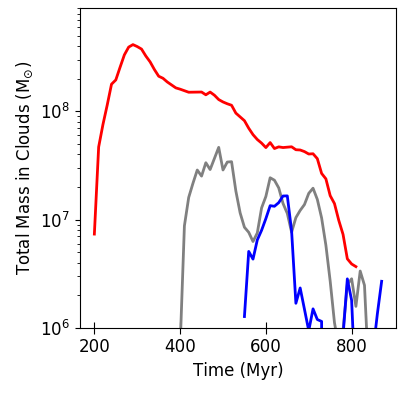}
\includegraphics[scale=0.57,trim= 0mm 2mm 0mm 2mm, clip]{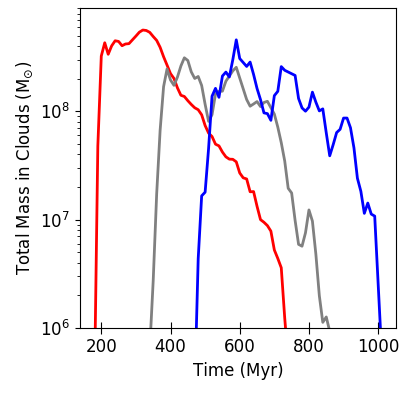}
\includegraphics[scale=0.57,trim= 0mm 2mm 0mm 2mm, clip]{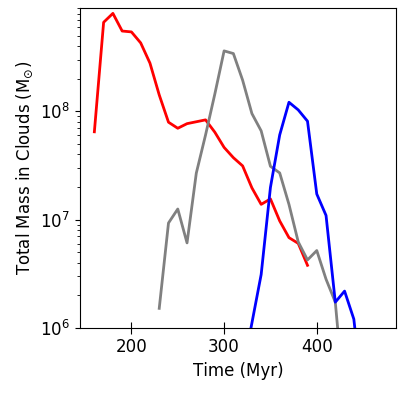}\\
\includegraphics[scale=0.57,trim= 0mm 2mm 0mm 2mm, clip]{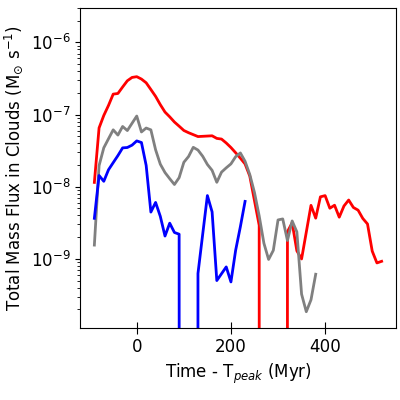}
\includegraphics[scale=0.57,trim= 0mm 2mm 0mm 2mm, clip]{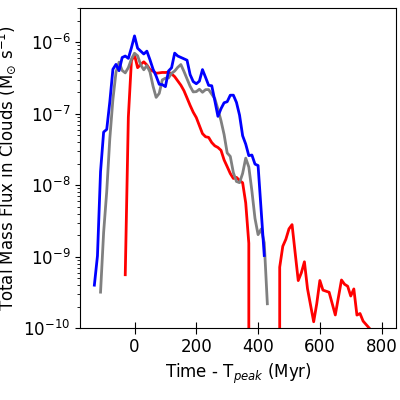}
\includegraphics[scale=0.57,trim= 0mm 2mm 0mm 2mm, clip]{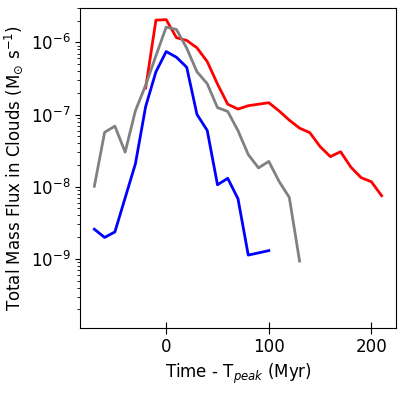}\\

\caption{The number of clouds (top panels) and the total mass in clouds (bottom panels) as a function of time in the three simulations, in each of the three height ranges previous defined.  Note that the x- and y-axis ranges differ across the panels.  The bottom panel shows the mass flux through each 6 kpc region as a function of time measured from the time of peak flux (Time - T$_{peak}$). Although the number and mass of clouds is highest in each simulation close to the disk, in HDLV the number of and mass in clouds may increase from $\sim$85 kpc to $\sim$155 kpc, and the mass flux increases as a function of height above the disk.}
\label{fig:numclouds}
\end{figure*}

We first focus on the shift of the simulated output off the model curve.  First, we note that this only occurs in the two LV runs near the galaxy disk.  As we have discussed with regards to Figure~\ref{fig:cartoon}, there are two main variables that change the shape of the velocity-ICM fraction curve in our model: the wind velocity and the galaxy escape velocity.  Including the galaxy escape velocity in our model results in a lower velocity at a given ICM fraction than we see in our simulations.  However, we note that even simply using a linear relationship due to momentum transfer would still predict velocities slightly below the peak of the HDLV contours.   

We next focus on our predictions for where along the velocity-ICM fraction curve gas should lie as a function of height above the disk.  We note that the LV runs show good agreement between the predictions and simulations, although the model has a slight shift towards higher velocities and ICM fractions as a function of height.  In the HDHV simulation, these predictions differ much more dramatically, with the difference between the simulation and model increasing with height above the disk. 

This could be for several reasons. First, we assumed a constant cloud mass, which is likely not the case -- clouds could be constantly accreting gas and fragmenting, processes which are not included in our simple model.  Indeed, as we discuss in the next section, we expect that not all clouds survive.  On the same note, we assume a constant 80 pc radius.  The clouds could be very elongated along the wind direction and be narrower than we model, which would shift our model shaded regions to lower velocities and ICM fractions.  Finally, for simplicity we assume that the ICM fraction of cloud gas changes linearly with time, but as the velocity difference between the cloud and ICM decreases, the mass deposition rate should also decrease, leading to a slower increase in the ICM fraction.  As we see in HDHV, this should lead to a shift in our model towards lower velocities and ICM fractions, particularly at large distances above the disk.  

\section{Cloud Survival}\label{sec:survival}

Beyond our model, the evolution of clouds in stripped tails is important for understanding gas mixing and cooling as well as star formation in the ICM.  As we have discussed in the Introduction, there remains debate about whether dense gas is removed from the disk and survives in the tail, or whether dense gas can form within that tail from lower-density galactic gas that is more easily stripped.

In this section, we directly address the evolution of clouds in our simulations, first empirically by examining cloud properties in the simulations as a function of height, and then theoretically by comparing the cooling and destruction (crushing) times of clouds.

\subsection{Cloud Number and Mass As a Function of Height Above the Disk}\label{sec:cloudsinsims}
\begin{figure}
    \centering
    \includegraphics[scale=0.5,trim= 0mm 2mm 0mm 1mm, clip]{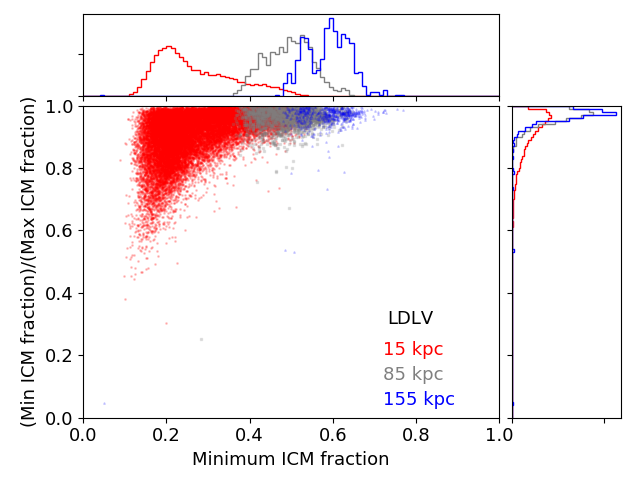}\\
    \includegraphics[scale=0.5,trim= 0mm 2mm 0mm 1mm, clip]{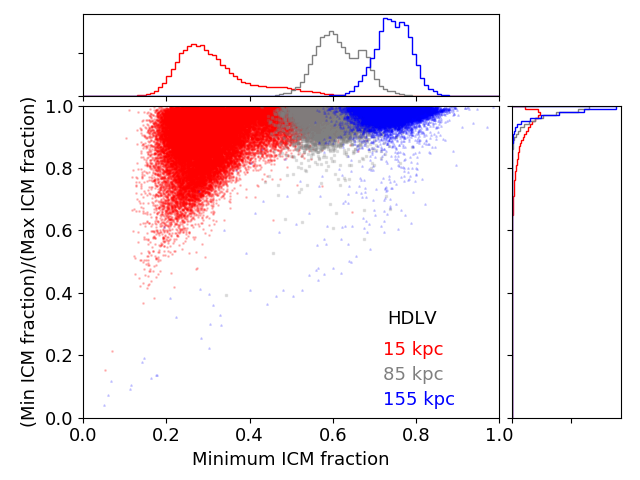}\\
    \includegraphics[scale=0.5,trim= 0mm 2mm 0mm 1mm, clip]{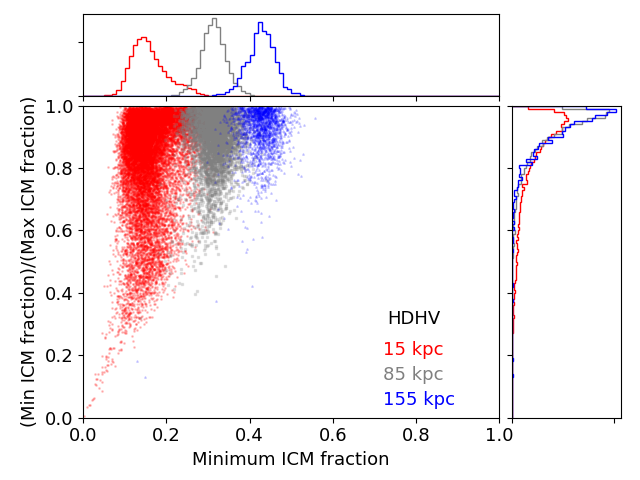}
    \caption{The ratio of the minimum to the maximum ICM fraction versus the minimum ICM fraction for clouds in the three simulations.  In all runs, the minimum ICM fraction increases as a function of height above the disk, indicating continued mixing of stripped and surrounding gas. Also, the ICM fraction within each cloud tends to be nearly uniform (and becomes increasingly uniform with height above the disk).}
    \label{fig:tracer_ratio}
\end{figure}

Because we cannot track individual clouds in our simulations, the most straightforward way to determine if clouds survive in the stripped tail is to simply count the number of clouds at different heights above the disk.  However, because clumps could fragment or merge, summing the total mass in clouds, or the mass flux, may be a more useful measure of the evolution of dense gas in the tail.  All of these metrics are shown in Figure~\ref{fig:numclouds}.  We note that clouds move at a range of velocities (as shown in Figure~\ref{fig:tracer_vz}), and so comparing gas at different heights is not likely to give us the exact same cloud population.  However, with that caveat in mind we make rough comparisons using the peak of the distributions at any height.

Clearly, in all simulations, the number of clouds decreases as we compare the 15 kpc height range to either 85 kpc or 155 kpc. Thus, many clouds are not surviving intact as they are being accelerated through the tail by the ICM wind. In agreement with this interpretation, we also see that the total mass in clouds decreases as we look farther from the disk.

However, when we compare the number of clouds at 85 kpc to that at 155 kpc, the fate of clouds becomes less clear.  In both the LDLV and HDHV simulations, the number of, and mass in, clouds at their peaks decrease from 85 kpc to 155 kpc.  In contrast, the peak number of clouds found at 155 kpc in the HDLV run is actually {\em larger} than the peak found at 85 kpc, and this increase in the number of clouds is reflected in an increase in the total mass in clouds.  While clouds may be both destroyed and formed in any of the simulated tails, the increase in cloud number and mass in the HDLV wind indicates that clouds are likely to be forming in the tail, with a net increase in the number of clouds beyond $\sim$85 kpc above the disk.

Because the cloud velocity varies as a function of height above the disk (and there is a range of cloud velocities at any given height as shown in Figure \ref{fig:tracer_vz}), we also look at the mass flux in clouds for a physically-motivated view of the cloud flow as a function of height. This is shown in the bottom panels of Figure~\ref{fig:numclouds}; here we have shifted the fluxes for the three different heights such that their peaks are coincident at $t=0$ in order to ease comparison of their amplitudes and widths. This confirms that the LDLV and HDHV runs have a lower flux of cold gas as we go away from the disk, while the HDLV run shows a growing flux with height. We will discuss a possible mechanism for this growth in Section~\ref{sec:wholewake}.

Using the number of clouds as a function of height, we have argued that many clouds are destroyed as they move away from the disk.  Next, we focus on the ICM fraction in order to determine whether clouds far from a galaxy originated in the disk or formed from the stripped gas.  As we showed in comparison to our model in Figure~\ref{fig:tracer_vz}, the mean ICM fraction in gas clouds increases with height above the disk in all runs, following our analytic framework that mixing is critical to acceleration.  In Figure~\ref{fig:tracer_ratio} we look more carefully at the distribution of the ICM fraction inside clouds by plotting the ratio of the minimum to the maximum ICM fraction of all the cells composing each cloud versus the minimum ICM fraction found in the cloud. We see that in general the ratio is close to one, meaning clouds are well-mixed.  Even though the scatter plot shows that the cells composing some clouds can have a range of ICM fractions, particularly in the HDHV run, the histograms demonstrate that the distribution peaks at high values (close to unity). 
\begin{figure*}
    \centering
    \includegraphics[scale=0.445,trim= 4.5mm 2mm 16mm 13mm, clip]{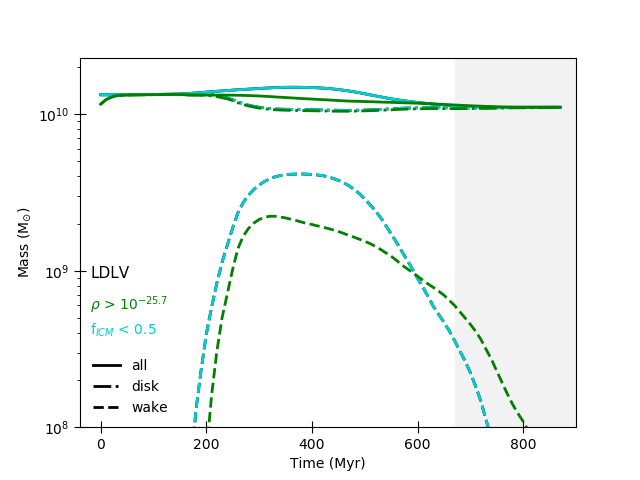}
    \includegraphics[scale=0.445,trim= 20mm 2mm 16mm 13mm, clip]{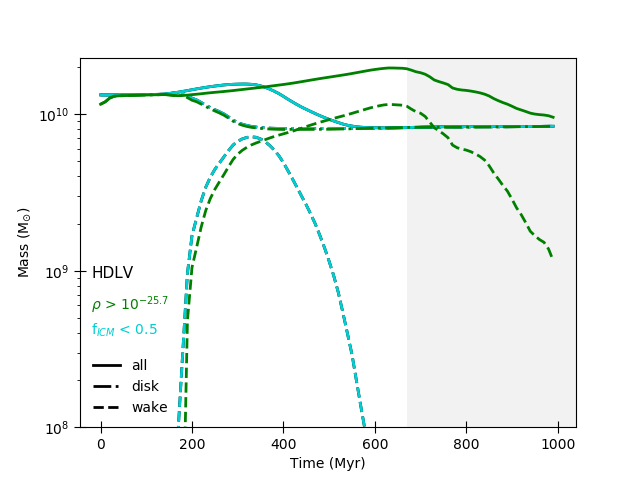}
    \includegraphics[scale=0.445,trim= 20mm 2mm 14mm 13mm, clip]{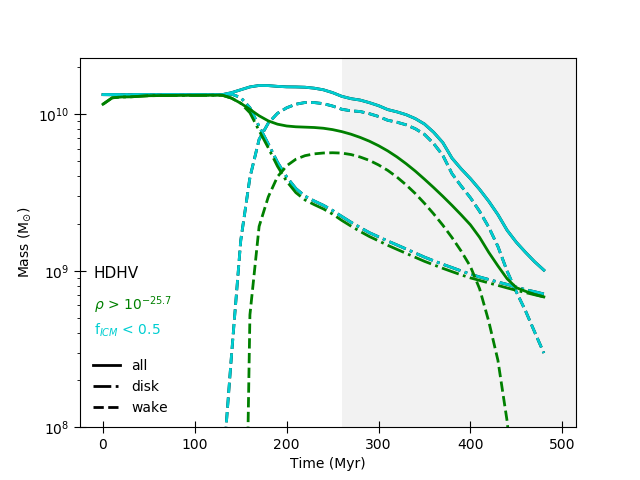}
    \caption{Mass versus time for gas in the entire box (``all"), within 10 kpc of the disk plane (``disk"), and from 10-224 kpc above the disk (``wake").  Each panel shows the gas mass for which $\ficm$ $<$ 0.5 (cyan) or $\rho$ $>$ 2$\times$10$^{-26}$ g cm$^{-3}$ (green).  In each simulation the mixing of disk gas with the ICM results in a different amount of dense gas relative to the low-$\ficm$ gas. Note that gas begins to leave the simulation volume after 670 Myr for the LV runs and 260 for the HDHV run and so care should be taken interpreting results after those times (shaded to guide the eye). The difference between the green and cyan dashed lines in the HDLV run signifies substantial condensation of ICM gas.}
    
    \label{fig:mass_time}
\end{figure*}

Importantly, we highlight that as we move farther from the galaxy, the minimum ICM fraction increases. Thus it is clear that mixing between the stripped and surrounding gas occurs throughout our clouds and no unmixed gas survives to large distances.  In fact, as we move farther from the galaxy, the ratio of the minimum to maximum ICM fraction moves closer to unity.  This is particularly notable in the case of HDLV, which shows an increase in the number and mass of clouds from $\sim$85 kpc to $\sim$155 kpc.  Even though the total mass of clouds is increasing, it is not due to stripped clouds simply accreting ICM gas.  In other words, in all three simulations there is no surviving core of pure ``galactic material" that is being accelerated while the outer layers of the clouds are mixing.  We note that this result is robust even when only selecting clouds with at least 300 cells, as shown in Appendix \ref{app:bigclouds}.

\subsection{The Wake as a Whole}\label{sec:wholewake}

As we have shown in the previous section, it is useful to compare cloud properties at different distances from the stripped galaxies.  However, we must be careful not to assume that clouds move together in clumps, or in other words that clouds found in the 15 kpc region at one output will later be found together 85 kpc from the disk.  This is visually clear when we consider the different distribution of clouds as a function of time in Figure \ref{fig:projections}, and is also predicted by the velocity range of clouds shown in Figure \ref{fig:tracer_vz}.

It is informative, then, to step back and measure the amount of dense and unmixed gas as a function of time in the simulations as a whole to see if they follow our expectations.  In Figure \ref{fig:mass_time} we do just that.  For each simulation, the mass of gas with $\ficm$ $<$ 0.5 is shown in cyan (i.e. gas originating primarily from the galaxy), and the mass of gas with $\rho$ $>$ 2$\times$10$^{-26}$ g cm$^{-3}$ is shown in green (i.e. dense gas of all origins).  The linestyles denote different spatial regions of the simulation. The early rapid increase in the dense gas (in the first 20 Myr of the simulation) comes from the relaxation of the disk, as dense clouds form out of the initially smooth disk gas.

Before exploring the ramifications of these plots, we note a technical point, which is the flow of gas out of the simulation volume. Because we are only showing here dense or low $\ficm$ gas, it is not clear whether decreasing mass corresponds to gas leaving the simulation volume, or transitioning into another state. However, we can resolve this difference by noting that in both LV runs the total amount of gas in the simulation decreases after about 670 Myr, and in HDHV the total amount of gas in the simulation decreases after about 260 Myr, indicating that at these late times more gas is leaving the box than is flowing into it (shown as the shaded regions in each panel).  This explains most or all of the late-time decrease in wake gas for both HD runs. 

If we first consider the mass of gas with $\ficm$ $<$ 0.5 (with a primarily galactic origin), we see that in all simulations, at early times (at $t \sim 200$ Myr for the LV runs and $t \sim 150$ Myr for the HV run), the total gas mass in the box increases as pure ICM gas is beginning to mix with the disk gas.  Comparing the wake to the disk gas masses, we see that most of the increase in gas mass is occurring behind the disk (in the wake).  This agrees well with our main prediction that mixing drives gas acceleration.  We also see a rapid drop in the wake mass as gas that is stripped continues to mix to high $\ficm$ values, as seen in Figure~\ref{fig:tracer_ratio}.

When we examine the dense gas in the simulation we see a somewhat different story. In all simulations, the amount of dense gas in the disk region decreases as the amount of dense gas in the wake increases, indicating that dense gas is stripped from the galaxy. However, the total amount of dense gas and the ratio between dense and low $\ficm$ gas differs dramatically in the different simulations.

In comparing the dense gas mass to the low $\ficm$ gas mass we will move from right to left in the panels.  In HDHV there is never an increase in the total dense gas mass after the wind hits the disk.  When we compare the wake gas mass measured using $\ficm$ or by our density cut, the HDHV curves are similar in width, but the amount of low $\ficm$ gas in the wake is higher than the amount of dense gas, indicating that in general mixing is heating stripped gas.

However, HDLV seems to tell the opposite story.  The total dense gas mass nearly doubles from the initial mass to the peak mass, and reaches much higher values than the total low $\ficm$ gas.  The dense gas peak in the wake is much broader and higher than the $\ficm$ peak, indicating that as gas mixes it continues to cool. This indicates that we are seeing significant gas condensation out of the ICM, driven by the mixing from clouds stripped out of the galaxy.

Finally, the LDLV run does not necessarily tell a single story of mixing.  As with HDHV, the total low $\ficm$ gas rises while the total dense gas falls, and in the wake the peak low $\ficm$ gas is higher than the peak dense gas. Interestingly, if we look closely at the wake gas in LDLV, the mass of dense gas has a slightly broader time profile than the mass of low $\ficm$, and there is more dense gas than low $\ficm$ gas in the wake after 600 Myr.  This may speak to late cooling of mixing gas, reminiscent of (but weaker than) the HDLV run.

To summarize this section and the previous one, we find that in all three simulations, $\ficm$ of clouds increases with distance from the disk, demonstrating mixing.  Importantly, this does not seem to correspond to universal cloud destruction or accretion. In more detail, as we move beyond 15 kpc from the disk, clouds seem to be destroyed in two simulations: the number and mass of dense gas in clouds continues to decrease throughout the wake and in time in both LDLV and HDHV.  However, looking farther in the wake in HDLV we see that the mass of clouds increases at large distances, and this is reflected in the total dense mass in the wake.

\begin{figure*}
\includegraphics[scale=0.37]{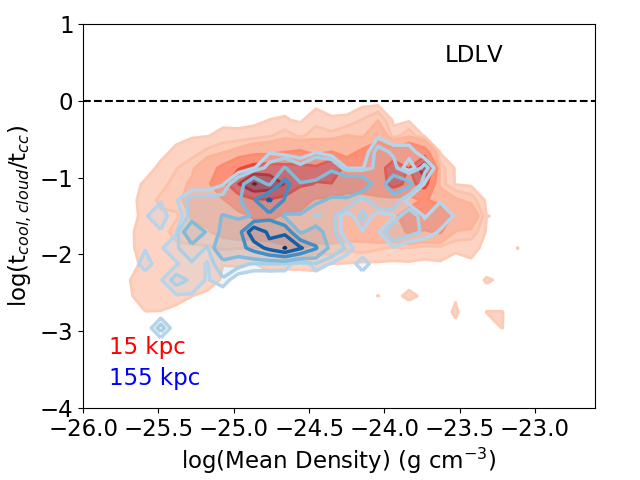}
\includegraphics[scale=0.37]{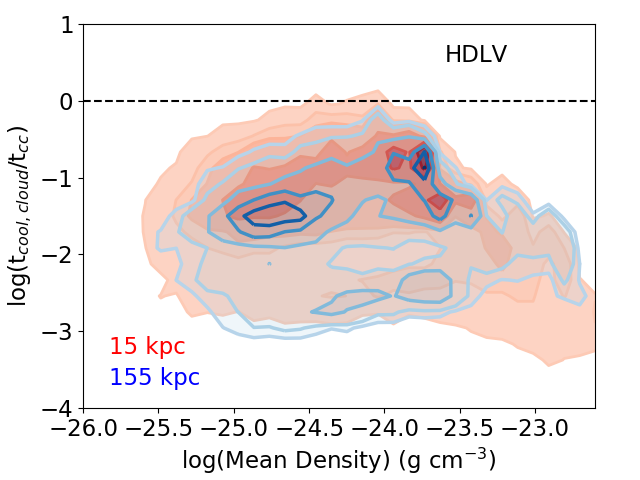}
\includegraphics[scale=0.37]{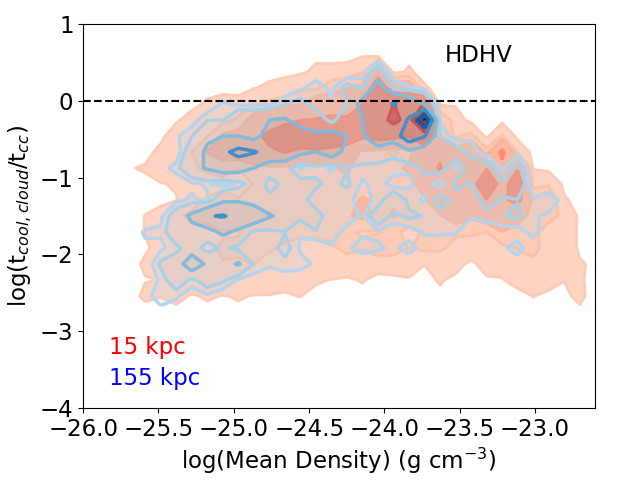}\\
\includegraphics[scale=0.37]{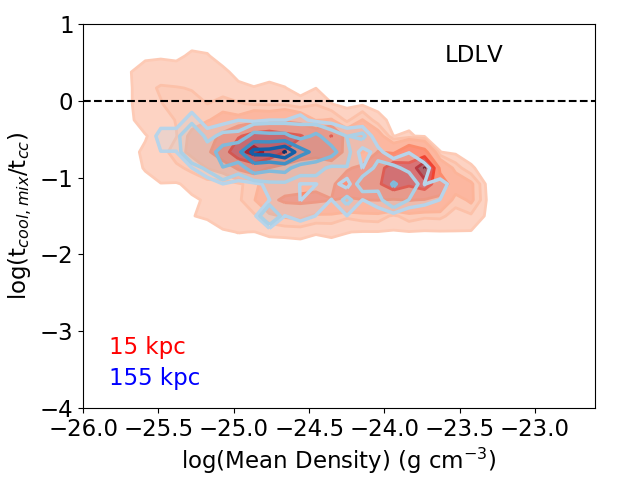}
\includegraphics[scale=0.37]{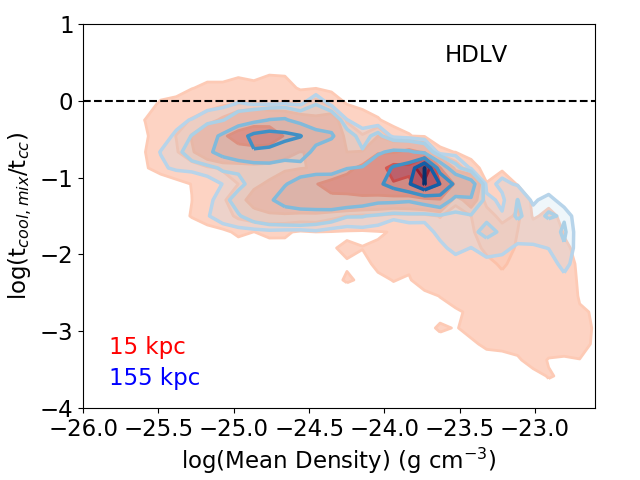}
\includegraphics[scale=0.37]{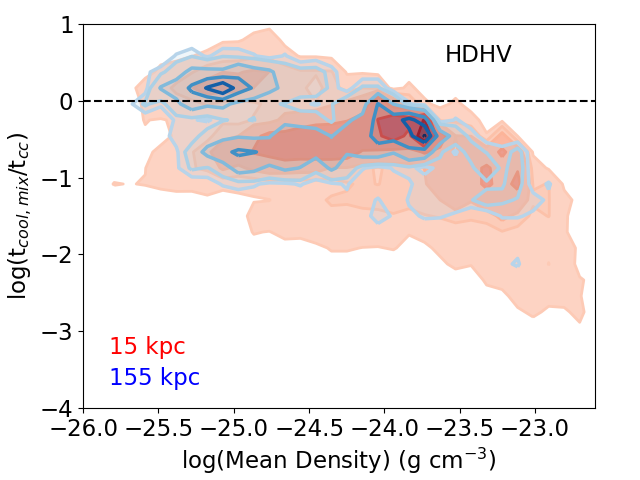}\\
\caption{The ratio of t$_{\rm cool}$/t$_{\rm cc}$ versus the mean density of the cloud for all three simulations.  As in Figure \ref{fig:tracer_vz}, the contours are based on the density of clouds in each two-dimensional histogram bin. In the top panels we use the maximum density and minimum temperature of the cloud to compute the cooling time, and generally find that these clouds should cool.  However, in the lower panels we use the mixing temperature and density as in Gronke \& Oh (2018) and find that these clouds lie much closer to the t$_{\rm cool}$ $=$ t$_{\rm cc}$ boundary (marked as a dashed line).  Comparing the nearest (red contours) to the most distant clouds (blue contours), we see that while the distributions of t$_{\rm cool}$/t$_{\rm cc}$ have a large amount of overlap, in both HD simulations clouds near the disk extend to much lower t$_{\rm cool, mix}$/t$_{\rm cc}$ ratios. }
\label{fig:coolingtimes}
\end{figure*}

\subsection{The Cooling versus Destruction time of Clouds}\label{sec:coolingtime}

We also consider the survival of clouds in the stripped tails more theoretically by comparing the cooling and destruction, or ``crushing" times of the clouds.  To calculate the cooling time of the gas we use the cooling rates from the grackle cooling tables implemented in our simulations.  Since, as we have shown, the metallicity varies little in the clouds, we use the average metallicity in the cooling calculation. 

We calculate the cloud crushing timescale as (e.g., Klein et al. 1994):

\begin{equation}\label{eqn:tcc}
    t_{cc} = \sqrt{\frac{\rho_{\rm cloud}}{\rho_{\rm ICM}}} \frac{R_{\rm cloud}}{v_{\rm diff}}
\end{equation}

For our calculation, $\rho_{\rm cloud}$ is the maximum cloud density, $\rho_{\rm ICM}$ is the inflowing ICM density, $v_{\rm diff}$ is the cloud velocity in the wind direction subtracted from the input wind velocity, and we assume spherical clouds and use the cloud volume to calculate $R_{\rm cloud}$.

In Figure~\ref{fig:coolingtimes} we plot the ratio of the cooling and crushing times as a function of mean cloud density. We calculate the cooling times of the clouds with two different assumptions.  In the top panels we use the minimum temperature of the cloud and the maximum density to calculate the maximum H number density using $\mu = 1$.  In the bottom panels we use the ``mixing" temperature and density (with $\mu = 0.6$) from Gronke \& Oh (2018; GO18). The distributions are similar, but we find that low density clouds are closer to t$_{\rm cool}$ $=$ t$_{\rm cc}$ using this second definition, and therefore are more likely to be destroyed before they cool.  This is largely because of the dependence of the cooling rate on density squared, which results in faster cooling rates for higher density gas.  Only our densest clouds have temperatures cooler than the peak of the cooling curve, so the rate is not generally strongly dependent on temperature.  Thus, while the cooling rate is quite similar for the cloud and the mixing layer, the cooling time is longer in the mixing layer because of the higher temperature. However, using either formalism we (marginally) find that most clouds will cool before they are destroyed. 

We also note that the ratio $t_{\rm cool}/t_{\rm cc}$ of clouds at different heights above the disk have very similar distributions.  However, in both the HDLV and HDHV simulations, clouds with higher densities extend to lower $t_{\rm cool}/t_{\rm cc}$ when they are close to the galaxy.  In addition, the fraction of clouds with $t_{\rm cool}/t_{\rm cc} > 1$ increases as we look farther from the disk in HDHV.

We highlight that HDLV has the least gas above the $t_{\rm cool} = t_{\rm cc}$ threshold, and extends to the most extreme (small) $t_{\rm cool}/t_{\rm cc}$ ratios.  The HDLV simulation has an ICM wind in which the number and mass of clouds increases from $\sim$85 kpc to $\sim$155 kpc, and in which the dense gas in the entire box increases over time. This is qualitatively consistent with the small $t_{\rm cool}/t_{\rm cc}$ ratios.   

How does this theoretical calculation align with our findings in Section~\ref{sec:cloudsinsims}, that clouds do not (generally) survive intact in the tail?  In the GO18 picture, even clouds that accrete gas first seem to lose dense gas mass, and only then cool and accrete gas in tails very late in the simulation (more than ten cloud crushing times at the highest density ratios).  Therefore, it is likely that the clouds that eventually accrete mass in a GO18 scenario are well-mixed throughout, consistent with our Figure~\ref{fig:tracer_ratio}. 

However, both the number and mass of clouds in our simulations decreases from 15 kpc above the disk to larger distances.  In GO18, higher density ratios between the surroundings and the cloud result in longer lag times before the dense cloud mass starts to increase.  In our ICM wind simulation, with higher density ratios than simulated in GO18 (or Gronke \& Oh 2019), the growth rate of these clouds may be even slower (comparing the ICM density to the average cloud densities from Figure \ref{fig:coolingtimes}, our clouds range from density ratios of $\sim$100 - 10$^4$).  The cloud crushing times range from 3$\times$10$^5$ years (the shortest crushing time in the HDHV run) to 10$^8$ years (the longest crushing time in HDLV).  However, the distribution of crushing times in HDHV peaks at 2.5$\times$10$^6$ years while both LV runs peak between 1-2$\times$10$^7$ years.  Even using the maximum cloud velocities from Figure \ref{fig:tracer_vz}, the travel time from 15 kpc to 155 kpc above the disk is more than ten cloud crushing times for the majority of clouds.    Therefore, only if the growth times of our clouds are more than ten cloud crushing times may we need to follow our clouds longer in order to see growth, or at density ratios of $\sim$10$^4$ clouds may not accrete gas from their surroundings.  

In all three simulations, as we have shown, our clouds straddle the $t_{\rm cool}= t_{\rm cc}$ line.  By using the maximum density for our clouds we are choosing the maximum possible crushing time for our comparison.  Further, we are using the mass density of the gas and an estimated $\mu$ as a proxy for H number density in our cooling time calculation.  These approximations could shift our clouds above or below the line of equality.  We are performing high resolution simulations of individual clouds to follow their evolution in detail and verify these calculations (Smith et al, in prep and Abruzzo et al., in prep).

\section{The Influence of the ICM on Cloud Properties}\label{sec:ICMinfluence}

In this paper we have used three simulations to support our model for ICM-mixing as the driver for gas acceleration in ram pressure stripped tails.  While we have discussed some differences in tail properties as a function of height above the disk and across the different simulations, in this section we focus on the impact of the ICM properties on cloud properties.  We find that both the ICM density and velocity influence the gas in the stripped tail.  

\subsection{ICM Density}

First, we compare the LDLV and HDLV simulations to discuss the effect of the ICM density on cloud properties.  In Figure~\ref{fig:tracer_vz} we showed that at the same height, the clouds in HDLV are moving slightly faster than those in LDLV, and in Figure \ref{fig:projections}, we see that HDLV has a longer dense tail than LDLV at the same simulation time.  In our model, this follows from the fact that there is more mass in the higher density wind resulting in quicker mixing for HDLV clouds.  

We also find that the maximum number and mass of clouds in any height bin is larger in the HDLV run than in the LDLV simulation.  While this must, to some extent, reflect the fact that the ram pressure ($\rho v^2$) is about twice as high in HDLV than in LDLV, and there is therefore more stripped gas that could cool to dense clumps, we argue that that is not the whole story. 

If we compare the relative number of clouds as a function of distance from the galaxy we see a dramatic difference in the two runs.  From Figure~\ref{fig:numclouds}, we can see that, at their respective peaks, there are about six times as many clouds at 15 kpc from the disk as 85 kpc above the disk in LDLV, while there are about two times as many clouds near the disk when doing the same comparison for the HDLV run.  The difference is even more dramatic when we compare the number of clouds at 155 kpc in the two LV runs.  The maximum number and mass of clouds in HDLV is larger at 155 kpc than at 85 kpc, in stark contrast to the continued decline in the number and mass in clouds as a function of height in LDLV.  

In addition, if we track the total dense gas mass within the entire simulation, as in Figure \ref{fig:mass_time}, we find that the dense gas mass increases in HDLV and decreases in LDLV.

This suggests that a higher density ICM leads to more cloud condensation due to stripped gas mixing.  This is consistent with the results of section~\ref{sec:coolingtime} -- in particular, when we look closely at Figure~\ref{fig:coolingtimes}, we see that LDLV has the largest population of clouds for which $t_{\rm cool,mix} \le 0.01 t_{\rm cc}$.

Examining the ICM fractions of dense gas in the HDLV and LDLV simulations can give us insight into whether mixing or another process such as compression is driving the differences in cloud mass in HDLV and LDLV.  Clouds in the HDLV tail tend to have higher ICM fractions, with the difference becoming more pronounced at larger height from the galaxy, as seen in both Figures~\ref{fig:tracer_vz} \& \ref{fig:tracer_ratio}.  This again follows our model as more mass from the wind will have impacted a cloud in a higher density ICM.  The different ICM densities do not have a dramatic impact on the minimum-to-maximum ICM fraction ratio, although there is a slight tendency for clouds in the LDLV run to have closer to uniform ICM fractions.  Therefore cloud mixing occurs in a similar fashion in both simulations.

Comparing the time evolution of low $\ficm$ gas to dense gas in Figure \ref{fig:mass_time} also indicates a significant difference in how mixing effects gas in the stripped tails due to ICM density.  In both runs the low $\ficm$ gas increases first, then as gas continues to mix with the ICM the low $\ficm$ gas mass decreases.  In LDLV the dense gas mass decreases when the low $\ficm$ gas decreases, indicating that mixing in low density ICM results in gas heating. On the other hand, in the higher density ICM of HDLV, low $\ficm$ gas mass decreases while dense gas mass continues to increase, indicating that mixing in HDLV results in gas condensing into clouds.

In summary, we argue then that stripped gas in a higher density surrounding medium accretes more of the ICM, resulting in gas with higher $\ficm$, \textit{and} the formation of more high-density clouds, particularly at larger distances from the galaxy.

\subsection{ICM velocity}

We can now compare HDLV and HDHV to discuss the effects of the ICM wind velocity on the stripped gas.  Unsurprisingly, the most clear difference is that stripped gas is moving faster in the high velocity run. 

Comparing the number and total mass of clouds as a function of height, we see that although the maximum number of clouds near the disk in HDHV is higher than in HDLV, as we would expect since the higher ram pressure in HDHV will strip more gas, the number drops by a slightly larger factor from 15 kpc to 85 kpc in HDHV than in HDLV at their peaks.  However, the difference is dramatic at the largest distance we probe--155 kpc from the disk, there are more clouds in HDLV than in HDHV at their peaks.  This indicates that the fast, high-Mach number HDHV wind does not allow dense clouds to survive (or form) as easily as HDLV\footnote{However, as we do not track individual clouds we cannot discount the possibility that there is a larger range of cloud velocity in HDHV that is spreading an increasing number of clouds more thinly across the length of the tail.}. Indeed, we also see this in the mass of dense gas in the wake as a function of time -- the mass drops dramatically in HDHV while it increases in HDLV (Figure \ref{fig:mass_time}).  This agrees with our finding in Figure~\ref{fig:coolingtimes} that more clouds have shorter $t_{\rm cc}$ than $t_{\rm cool}$.

Interestingly, the gas composing HDHV clouds has a large range of ICM fractions, particularly close to the galaxy (Figure~\ref{fig:tracer_ratio}).  Perhaps the range in ICM fractions within a cloud is a signature of the clouds being destroyed, as indicated by the short cloud crushing times.  We also find that at the same ICM density the faster wind (HDHV) results in clouds with lower minimum ICM fractions (in other words, less mixing between the stripped and surrounding gas).  This supports our analytic model that the faster ICM will add more momentum to stripped gas per unit mass.  In addition, almost all identified clouds in HDHV have minimum $\ficm$ less than 0.5.  In combination with the fact that the mass of low $\ficm$ gas in the wake is always more than the mass of high density gas (Figure \ref{fig:mass_time}), this again highlights that mixing is likely to heat cold, dense gas into the diffuse ICM.

\section{Discussion}\label{sec:discussion}

\subsection{The Importance of the ICM}

In previous work we have argued that the ICM is important in setting the properties in the tail, from the X-ray luminosity to the star formation rate (Tonnesen et al. 2011; Tonnesen \& Bryan 2012).  We found that both the X-ray luminosity of the stripped tail and the star formation rate in stripped gas will increase as the ICM pressure increases due to the near-pressure equilibrium of the stripped and surrounding gas.  These conclusions were largely based on the $\rho$-T diagrams of the stripped gas, which found higher density gas at all temperatures in a higher-pressure ICM. 

In this paper, we focus on the properties of overdense clumps in the tail. As we have discussed, we find agreement with our previous conclusion that a higher density ICM produces more high density clouds.

Here we look more closely at our clouds to determine whether they are likely to be gravitationally bound or pressure confined.  To do this we compare estimates of the cloud internal energy and the gravitational potential:
\begin{equation}
    U = \frac{3}{2}k_{\rm boltz}T_{\rm mean}\frac{M_{\rm cloud}}{\mu m_{H}}
\end{equation}
\begin{equation}
    U_g = \frac{GM_{\rm cloud}^2}{r_{\rm cloud}}
\end{equation}
In which T$_{\rm mean}$ is the mass-weighted average temperature, M$_{\rm cloud}$ is the total mass of the cloud, and r$_{\rm cloud}$ is the cloud radius assuming a spherical cloud with the volume of cells included in the identified cloud.  We are ignoring turbulent energy and assuming all internal energy is thermal.

\begin{figure}
    \centering
    \includegraphics[scale=0.52,trim=0mm 15.2mm 0mm 5mm,clip]{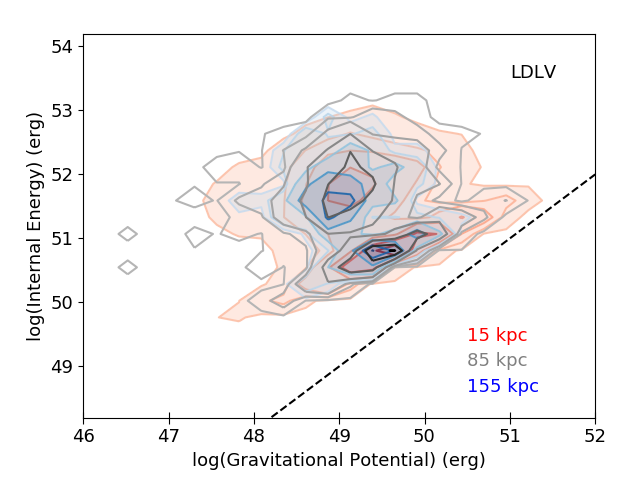}\\
    \includegraphics[scale=0.52,trim=0mm 15.2mm 0mm 8mm,clip]{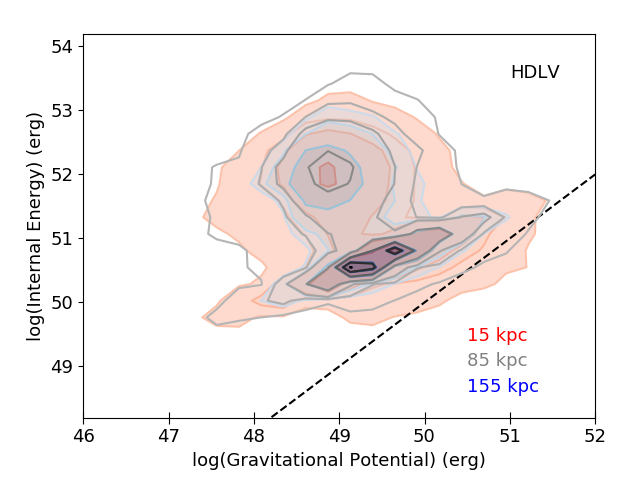}\\
    \includegraphics[scale=0.52,trim=0mm 1mm 0mm 8mm,clip]{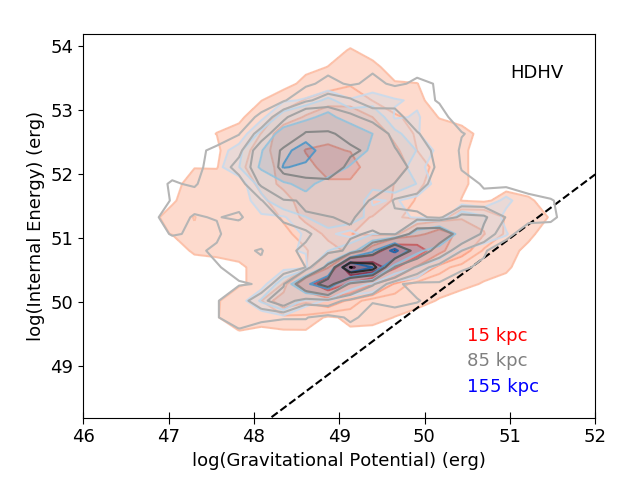}
    \caption{The internal energy of clouds versus the gravitational potential of clouds.  As in Figure \ref{fig:tracer_vz}, the contours are based on the density of clouds in each two-dimensional histogram bin. The panels are organized as in Figure \ref{fig:tracer_vz}, and have a line at $x=y$.  We see that the internal energy of the clouds is higher than the gravitational potential, indicating that these clouds are pressure confined.}
    \label{fig:cloudconfinement}
\end{figure}

Figure~\ref{fig:cloudconfinement} shows the distribution of these two quantities for the clouds in all three simulations at all three heights.  From this, we see that most of the clouds in all three simulations are pressure confined ($U_g \ll U$).  In our higher-density ICM runs (HDLV and HDHV), some of the most massive clouds lie very close to the line of equality.  These clouds seem to be largely pressure confined, which agrees with our earlier findings of near-pressure equilibrium in the stripped tail and ICM, although we cannot rule out a part of these clouds being gravitationally unstable. Indeed, when we select only clumps with at least 300 cells, as in Appendix \ref{fig:appendix_sel300}, we find a much higher fraction of massive clouds lie close to the line of equality, supporting the idea that regions of massive clouds may be gravitationally unstable.

Because our model for gas acceleration requires the mixing of ICM and ISM gas, we have carefully examined the mixed fraction of dense clumps.  In Figure~\ref{fig:tracer_ratio} we saw evidence that the surrounding ICM is not just compressing the stripped material towards pressure equilibrium, but is mixed throughout the stripped gas, even in the most dense clouds.  In fact, we see that this mixing is more efficient in the HDLV run than the LDLV run, indicating that a higher density (and therefore higher pressure) ICM mixes more with the stripped tail.  

The mixing of the stripped and surrounding gas can either result in the cooling of hot gas or the heating of cold gas.  Clouds tend to lie close to the $t_{\rm cool,mix} = t_{\rm cc}$ line, indicating that they may be able to survive for significant amounts of time.  In addition, as we have mentioned in Section~\ref{sec:coolingtime}, the density ratio of clouds to the ICM ranges from 10$^3$ - 10$^5$.  Therefore, in a picture in which the mass accretion of dense clouds takes more time for higher density ratios as in GO18 (at fixed $t_{\rm cool,mix}/t_{\rm cc}$), the cooling onto these clouds could take more than 1 Gyr (10 cloud crushing times of 10$^8$ years).

These results lead us to a picture in which stripped gas quickly mixes with the ICM.  This happens more quickly when the surrounding gas has a higher density (and therefore pressure).  We posit that this plays out in different ways in the HDLV and HDHV runs.  In pressure equilibrium, cold gas will have higher densities so the cooling time will be shorter and clouds will be more likely to survive.  We see this in the HDLV run in the low $t_{\rm cool,mix}/t_{\rm cc}$ clouds in Figure~\ref{fig:coolingtimes} and in the increasing number and mass of clouds from $\sim$85 kpc to $\sim$155 kpc.  However, although the cloud densities are similar in the HDHV run, the velocity difference between the cloud and the ICM is also larger, leading to a lower $t_{\rm cc}$, and therefore more clouds reside above the $t_{\rm cool,mix} $=$ t_{\rm cc}$ line in Figure~\ref{fig:coolingtimes}.  Therefore, when stripped gas has mixed with a large fraction of ICM gas (i.e. $\ficm \gtrsim$ 0.5), these clouds are destroyed (Figure \ref{fig:mass_time}).  This is why we do not see clouds with high ICM fractions in Figure~\ref{fig:tracer_ratio}.

Therefore the interplay of ICM density and velocity will determine whether gas mixing results in dense clouds or diffuse gas.

\subsection{Making Predictions for observations}\label{sec:observations}

We make a clear prediction that gas that is moving faster will be more highly mixed with the surrounding ICM, which should be reflected in the metallicity of stripped gas.  This may, in fact, be the case in stripped galaxies, for example in the system JO201 (Bellhouse et al. 2019).  However, in order to use this model to determine the three dimensional distance and velocity of  observed clouds, we would need to fold in the metallicity of the ISM as a function of the radius from which it was stripped.  As this becomes more well-constrained, our predictive power will increase.

We have also found that for galaxies moving at the same velocity with regards to the ICM, a higher density ICM will lead to tails moving at higher velocities and to more dense clouds at large distances from the galaxy.  We also argue that at the same ICM density, dense clouds are less likely to be found at large distances in tails stripped from galaxies with high velocities.  Although this is a relative relation, it is possible that it could be used to compare the likely ICM conditions surrounding galaxies undergoing ram pressure stripping and help to break degeneracies resulting from only observing the line of sight velocity and projected distance from the cluster center. 

\subsection{Star Formation in Stripped Tails}

Although we do not include star formation in these simulations and only focus on dense gas, in observations of stripped tails some of the best evidence for cold dense clouds is the HII regions resulting from star formation.  However, currently there are no clear cases of stars formed more than $\sim$100 kpc from a stripped disk (Poggianti et al. 2019), yet we find dense clouds out to at least $\sim$155 kpc.  

There are several possible explanations for this.  On the observational side, stars at such a large distance from the disk may not be as clearly associated with a stripped tail because all lower density gas will have been mixed and heated until it is indistinguishable from the ICM.  

On the simulation side, we first recall that our clouds are pressure confined, and our rough estimate finds that $t_{cool}\sim t_{cc}$, so many of our clouds could be destroyed before they would cool and form stars.  Also, it is possible that star formation in dense gas closer to the disk will heat nearby gas and cause it to mix into the surrounding medium more efficiently than we find in our simulations.  A highly-resolved simulation including star formation is required to determine how far from the disk stars will form.

However, despite these caveats, based on our results we would predict that star formation is most likely to occur in the stripped tails of galaxies moving slowly in a high-density ICM.  This might translate into more star formation from stripped galaxies with circular orbits close to the cluster center.

\subsection{Caveats}\label{sec:caveats}

In this section, we briefly discuss a number of shortcomings of our simulations, first discussing the impact of spatial resolution, before turning to other physical effects that we have not included.

\subsubsection{Resolution}

Although our resolution of up to $\sim$40 pc allows for many cells across the galaxy disk, individual clouds in the tail are not well-resolved.  We require at least 10 cells in a cloud, but in order to begin to resolve an individual cloud, $\sim$10$^5$ would be required (32 cells across the cloud radius).  
However, the individual clouds we identify are objects in the flow and have already been processed by turbulence and cooling, while the resolution requirement of clouds in cloud-crushing simulations refers to the initial cloud size -- such simulations also generally find structure in their partially mixed clouds down to the grid scale. Therefore, we argue that the in-situ cloud sizes are not necessarily a good estimate of the resolved nature of the simulations.

We examine the impact of resolution in more detail in Appendices A and B, where we look at the properties of clouds as a function of their resolution as well as an additional set of simulations performed at lower resolution. The overall conclusion is that, while detailed properties do change, the overall results are remarkably robust to changes in resolution.

However, we note that if a cloud does fragment into small pieces, as in the ``shattering" scenario of McCourt et al. (2018), because we are not resolving the $c_s t_{\rm cool}$ length scale, the fragments would mix into the surrounding medium at the grid scale.  
Our clouds are not in precise pressure balance with the ICM, perhaps making a shattering scenario more relevant (see discussion in Gronke \& Oh 2020).

Future simulations of clouds surrounded by gas at these high density and temperature ratios are required to predict the ICM densities and temperatures at which clouds will mix into the ICM or will accrete gas from the ICM.  In this paper we only claim that higher ICM density is likely to result in longer cloud survival, while higher relative ICM velocity will result in faster cloud mixing.

\subsubsection{Missing Physics}

We run these simulations using only hydrodynamic equations, which means that we may be missing relevant physics for gas mixing.   

For example, we have no magnetic fields.  Gronke \& Oh (2020) and others (e.g. Sur et al. 2014; McCourt et al. 2015; Cottle et al. 2020) include magnetic fields in their simulations that study the growth of cold clouds through the entrainment of the surrounding material.  Their magnetic fields increase the velocity of clouds, suppress the KH instability, and stretch the cold gas along the field lines.  Importantly, they find that magnetic fields have little impact on the overall mass growth rate.  In contrast, in plane-parallel simulations, Ji et al. (2019) find that magnetic fields suppress cold gas growth through entrainment.  While it is not clear whether magnetic fields would suppress or enhance the growth of cold clouds, it is likely that they would stabilize them against mixing to some degree.

Along with a magnetic field, we do not include thermal conduction.  Cloud survival has been studied including isotropic heat conduction.  Bruggen \& Scannapieco (2016) include both radiative cooling and heat conduction in simulations of clouds being ejected in galactic outflows.  They find that while the outer envelope is evaporated, the central region cools and stretches into dense filaments (in agreement with simulations that only include radiative cooling: Mellema et al. 2002; Fragile et al. 2004; Orlando et al. 2005; Johansson \& Ziegler 2013).  These clouds are accelerated only at early times because their cross-section decreases dramatically as the outer layers evaporate, so they move more slowly than simulated clouds without heat conduction. Clouds with heat conduction also lose mass more quickly than those with only radiative cooling.  

The authors do highlight that this time can be lengthened if there are magnetic fields perpendicular to the temperature gradient (Cowie \& McKee 1977; Cox 1979). Vollmer et al. (2001) argue that magnetic fields could increase the evaporation time by nearly an order of magnitude, although the actual value is highly uncertain. In previous work, we have found that in order for the length of tails in our simulations to agree with observations, heat conduction must not be efficient (Tonnesen \& Bryan 2010; Tonnesen et al. 2011).

As with increased resolution, we think that including these physical effects will not alter the trends we have found in this paper.  Interestingly, recall that in the HDHV run, the cloud velocities far from the disk are below our analytic prediction.  This may be explained by the single-cloud results that with radiative cooling clouds stretch into narrower filaments over time, while our analytic model assumes a constant cloud cross-sectional area.

\section{Conclusions}\label{sec:conclusion}

We have presented an analytic model describing how gas is unbound and accelerated away from a galaxy due to mixing-mediated ram pressure stripping. The model is based on the idea of mixing driving momentum transfer, rather than a traditional ram ``force".  To verify this model and to highlight the impact of the ICM velocity and density, we also present three ``wind-tunnel" simulations of a galaxy undergoing ram pressure stripping in which we identify and examine clouds in the tail.  Our main conclusions are as follows:
\begin{enumerate}
\item We present a model in which the acceleration of gas from a galaxy is due to mixing with the ICM, and is based on a mix of energy and momentum deposition into galactic gas from the ICM (Section~\ref{sec:mixing}).  The model makes clear predictions that gas with higher velocities and at larger distances from the disk will be more well-mixed with the ICM (Figures~\ref{fig:cartoon} \& \ref{fig:illustration}).  

\item We compare our model to three simulations in which we have varied the ICM velocity and density.  We find excellent agreement with the ICM fraction-velocity relation for all gas in the stripped tail (Section~\ref{sec:allgascomparison}).

\item When we focus only on the dense clouds in our simulations we still find good agreement, even when we fold in a model for the distance a cloud has traveled (Figure~\ref{fig:tracer_vz}).

\item Clouds are nearly uniformly mixed with the ICM, meaning that in our simulations there is no cloud ``core" that survives intact from the galaxy.  The ICM fraction in dense clouds increases as a function of height from the disk (Figure \ref{fig:tracer_ratio}).  Comparing simulations, we find that at the same wind velocity, clouds are more well-mixed in a higher density ICM.  At the same ICM density, clouds are more well-mixed in a slower wind. 

\item Both the number and mass of gas in dense clouds decreases as a function of height above the disk (Figure~\ref{fig:numclouds}), suggesting that clouds generally do not survive in the tail. However, in the HDLV run, the number and mass of clouds increases from $\sim$85 kpc to $\sim$155 kpc above the disk, indicating that some clouds survive and accrete gas from their surroundings. Stripped gas mixing with the ICM results in decreasing dense gas mass in HDHV and LDLV, but in HDLV mixing adds dense gas mass to the simulation  (Figure~\ref{fig:mass_time}).

\end{enumerate}

Importantly, we have shown in this paper that dense clouds in stripped tails are part of the continuum of gas in the tails.  Our mixing-driven model for gas acceleration in the tail applies equally well to low-density and high-dense gas.  This is an important departure from the simple picture that intact clouds can survive being ``pushed" from the galaxy by an ICM wind.  Indeed, this leads to the observational prediction that the metallicity of dense clouds should decrease as the distance from the galaxy increases (absent enrichment due to additional star formation in the tail).

Because mixing drives the formation and acceleration of the stripped gas, the question to ask when comparing the different temperature and density distributions of tails becomes clear: will mixing with the ICM result in more gas being able to cool into dense clouds, or will it heat the stripped gas until it is indistinguishable from its surroundings?  We have shown that a higher density ICM will tip the balance towards more dense cloud formation, while higher relative ICM velocities will destroy clouds and add the stripped gas to the diffuse surroundings.

Our results on cloud survival and mixing are based on comparing simulations with different ICM properties, however we also consider cloud survival more theoretically by calculating their crushing and cooling times.  We find that clouds fall near the $t_{\rm cool,mix} = t_{\rm cc}$ line when we use the cooling time of the mixing layer of our clouds (Figure~\ref{fig:coolingtimes}), indicating that cloud survival cannot be universally predicted even within a single stripped tail (or that the process of cloud evolution drives clouds close to this relation).  

Given the diversity of observed stripped tails, some with dense gas and even star formation, while others only contain more diffuse gas in their tails, the question of what causes dense gas survival and collapse to stars is important to understanding the physics at play in the ICM.  While our results are an important step outlining the competition between velocity and density, predicting the survival of dense gas in stripped tails in detail will require more theoretical work -- in particular, a suite of simulations that resolve individual clouds and include star formation as well as non-ideal MHD processes such as conduction.

\acknowledgements

We would like to thank the referee for helpful comments that improved the paper.  The authors gratefully acknowledge  support from the Center for Computational Astrophysics at the Flatiron  Institute,  which  is  supported  by  the  Simons Foundation.  ST thanks the GASP collaboration for useful discussions about gas acceleration in stripped tails.  GLB  acknowledges  financial  support  from   NSF   (grant   AST-1615955,   OAC-1835509),   and NASA (grant NNX15AB20G), and computing support from  NSF  XSEDE.  The  simulations  used  in  this  work  were run on facilities supported by the Scientific Computing Core at the Flatiron Institute, a division of the Simons Foundation.

\appendix
\restartappendixnumbering

\section{Clump Resolution}\label{app:bigclouds}

In this appendix we discuss individual cloud resolution by comparing the properties of clumps that include at least 300 cells to our entire set, which includes clumps with as few as 10 cells.   

\begin{figure}
    \centering
    \includegraphics[scale=0.36]{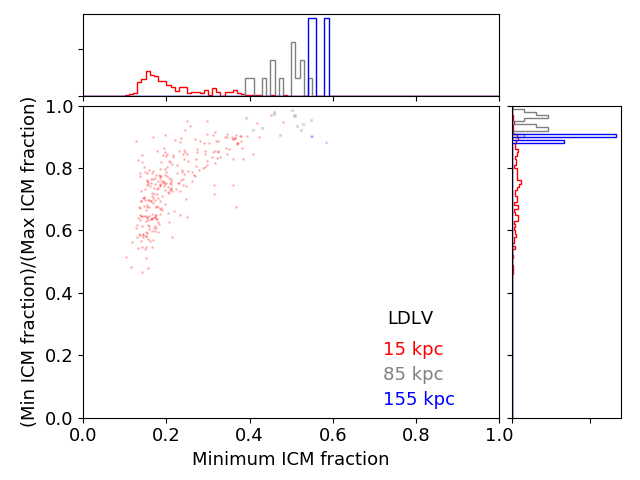} 
    \includegraphics[scale=0.36]{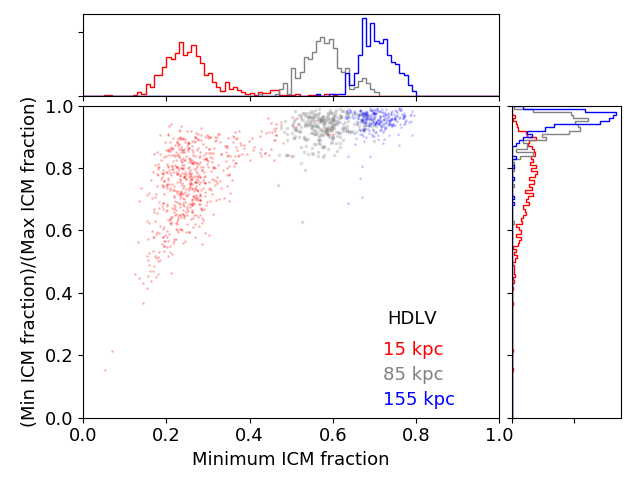}
    \includegraphics[scale=0.36]{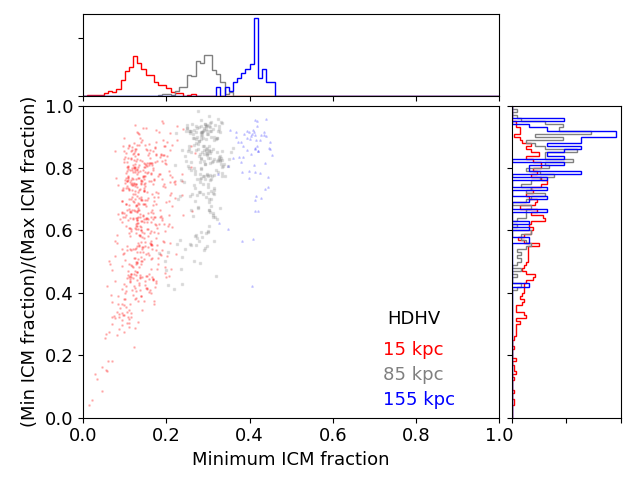}
    \caption{The same as Figure \ref{fig:tracer_ratio}, but only including clumps with at least 300 cells.  The same trends hold as in the full clump sample, and we can clearly see that individual clumps tend to be more uniformly mixed (have minimum-to-maximum ratios closer to unity) farther from the disk.}
    \label{fig:appendix_sel300}
\end{figure}

In Figure \ref{fig:appendix_sel300} we repeat Figure \ref{fig:tracer_ratio} using only those clumps with at least 300 cells.  The most striking, although unsurprising, difference is that there are many fewer clumps using this selection criteria.  We also note that the ratio between the minimum and maximum ICM fractions within each clump tends towards lower values.  Again, because we select larger clumps we would expect this result.  Interestingly, the differences between the ICM fraction minimum-to-maximum ratio as a function of height above the disk are clearer when we focus on larger clouds.  Specifically, clumps identified at 15 kpc above the disk tend to have lower ratios, in other words be less well-mixed, than those at 85 kpc or 155 kpc.  This agrees well with our conclusions that mixing is driving cloud acceleration and that unmixed cores are not surviving to large distances.  

\begin{figure}
    \centering
    
    \includegraphics[scale=0.36]{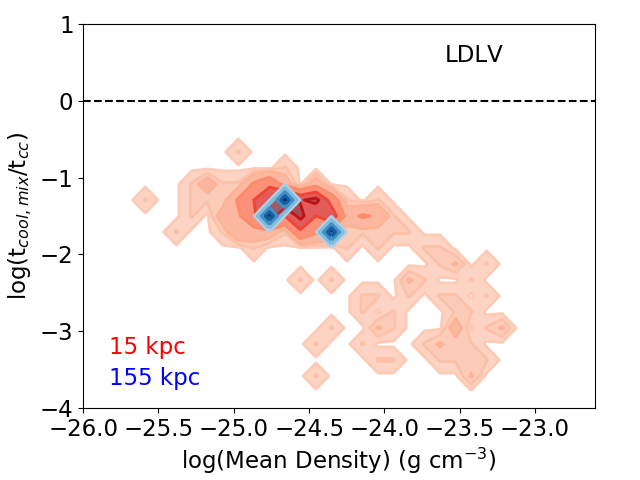}
    \includegraphics[scale=0.36]{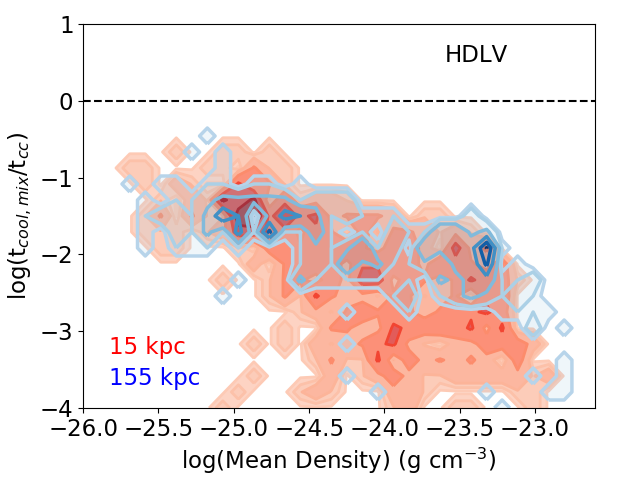}
    \includegraphics[scale=0.36]{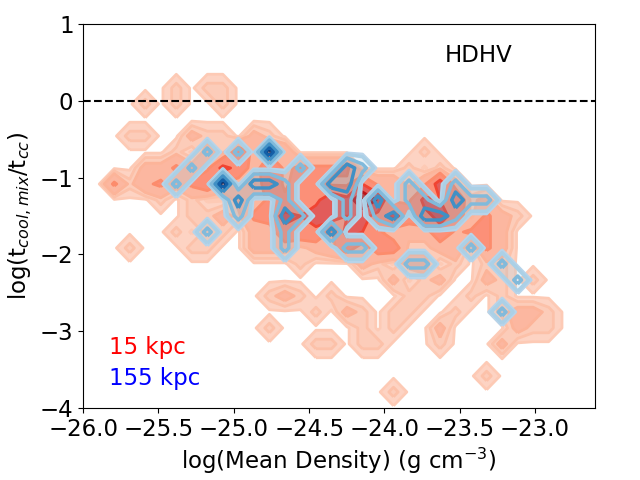}
    \caption{The same as the bottom three panels in Figure \ref{fig:coolingtimes}, but only including clumps with at least 300 cells.  These larger clumps tend to lie well below the line of equality.  }
    \label{fig:appendix_tcool}

\end{figure}

In Figure \ref{fig:appendix_tcool} we recreate the bottom panels of Figure \ref{fig:coolingtimes} to determine whether the trends we see for the whole cloud sample continue when only considering the largest clouds.  As one would expect, when focusing on larger clouds the $t_{\rm cool,mix}/t_{\rm cc}$ decreases.  As t$_{\rm cc}$ increases with cloud radius, even at the same cloud density we would expect a lower $t_{\rm cool,mix}/t_{\rm cc}$, as we see.  

When we look at only the largest clumps in our tails, we see that the vast majority lie well below the line of equality, indicating they will grow in mass rather than destroyed.  However, we note that the number, mass, and mass flux of large clouds decreases from 15 kpc to 155 kpc in both LDLV and HDHV.  

We also note that there still does seem to be a difference in the $t_{\rm cool,mix}/t_{\rm cc}$ distributions in the large clouds, with HDLV values tending to be lower.

\section{Simulation Resolution}

\begin{figure}
    \centering
    \includegraphics[scale=0.36]{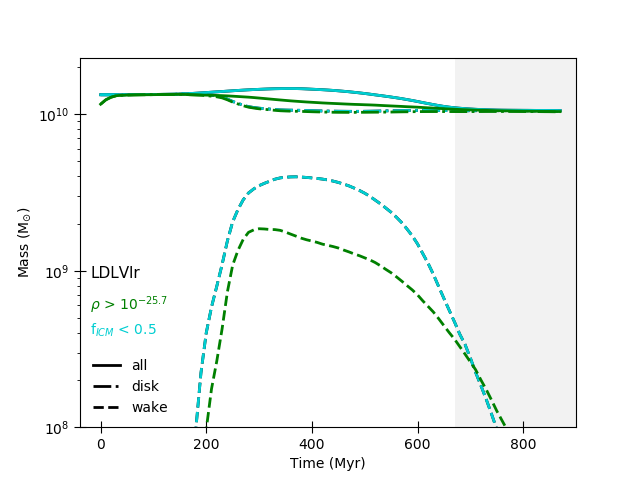} 
    \includegraphics[scale=0.36]{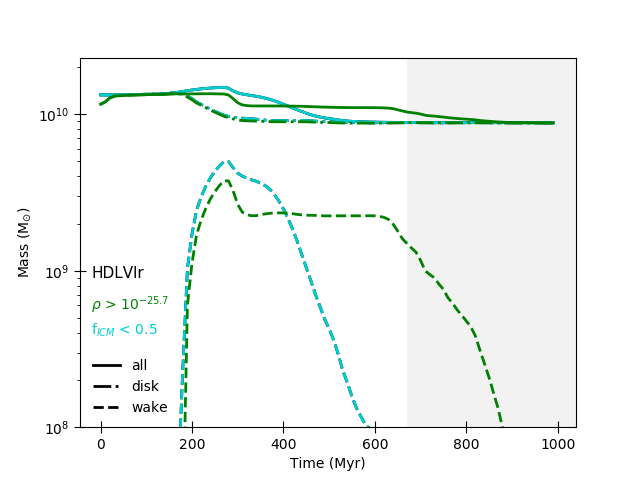}
    \includegraphics[scale=0.36]{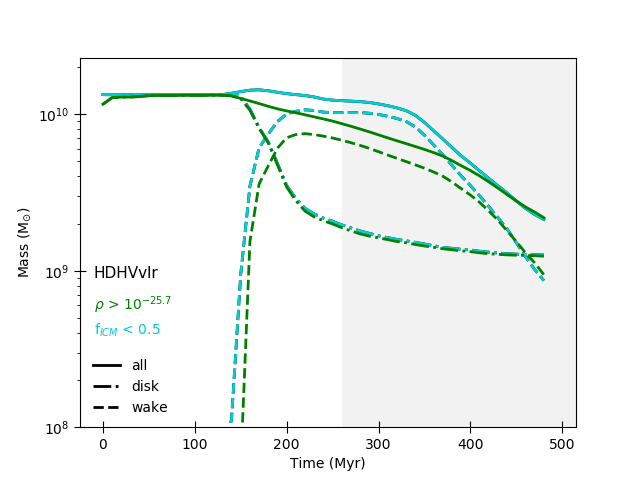}
    \caption{As Figure \ref{fig:mass_time}, but for simulations run at either 160 pc (LDLV and HDLV) or 320 pc (HDHV) resolution.  We see that particularly for LDLV and HDHV the unmixed and dense gas trends are remarkably similar to the 40 pc resolution runs.  Although the specifics of HDLV differ at the lower resolution, the fact that more dense gas than unmixed gas exists in the wake at late times is consistent with the higher resolution results.}
    \label{fig:appendix_masstime}
\end{figure}

In this section, we examine the impact of the maximum resolution in the simulations on the evolution of mixed and dense gas in the tails.  We have rerun all three simulations using lower resolution.  LDLV and HDLV were run with a maximum resolution of 160 pc, and HDLV was run with a maximum resolution of 320 pc.  In Figure \ref{fig:appendix_masstime}, we repeat Figure \ref{fig:mass_time} with the lower resolution simulations.

First, we highlight the striking similarity in the LDLV and HDHV simulations at high (40 pc) and low (160 pc or 320 pc) resolution.  At lower resolution, we see that there is always more low f$_{ICM}$ gas in the simulation than there is dense gas, and this is reflected in the wake.  This indicates that in general, as gas mixes it is heated and diffuses into the ICM.  However, as with the higher resolution runs, HDLV tells a different story.  Here the mass of dense gas in the wake is larger than the mass of low f$_{ICM}$ gas, indicating that gas mixes and cools.  Interestingly, we do not see the growth of high-density gas in the wake as in the higher resolution simulation.  We highlight that at lower resolution we find the same qualitative trends--mixing gas does not become dense in LDLV and HDHV, while mixing gas can become dense in HDLV.

\end{document}